\documentclass[twocolumn, twocolappendix]{aastex631}
\usepackage{CJKutf8}
\usepackage{amsmath}
\usepackage{amssymb}
\usepackage{comment}

\newcommand{\fehi}{[Fe/H]$_\mathrm{init}$}
\newcommand{\afei}{$[\alpha/\mathrm{Fe}]_\mathrm{init}$}
\newcommand{\Teff}{T$_\mathrm{eff}$}
\newcommand{\logg}{$\log{(g)}$}
\newcommand{\feh}{[Fe/H]}
\newcommand{\afe}{[$\alpha$/Fe]}
\newcommand{\Av}{$A_V$}

\newcommand{\logR}{$\log{(\mathrm{R})}$}

\newcommand{\angstrom}{\textup{~\AA}}

\shorttitle{Rapid Formation of the Metal Poor Milky Way}
\shortauthors{Woody et al.}

\graphicspath{{./}}
\begin{document}
\begin{CJK*}{UTF8}{gbsn}

\title{The Rapid Formation of the Metal Poor Milky Way}

\correspondingauthor{Turner Woody}
\email{turner.woody@cfa.harvard.edu}

\author[0000-0002-0721-6715]{Turner Woody}
\affiliation{Center for Astrophysics $|$ Harvard \& Smithsonian, 60 Garden Street, Cambridge, MA 02138, USA}

\author[0000-0002-1590-8551]{Charlie Conroy}
\affiliation{Center for Astrophysics $|$ Harvard \& Smithsonian, 60 Garden Street, Cambridge, MA 02138, USA}

\author[0000-0002-1617-8917]{Phillip Cargile}
\affiliation{Center for Astrophysics $|$ Harvard \& Smithsonian, 60 Garden Street, Cambridge, MA 02138, USA}

\author[0000-0002-7846-9787]{Ana Bonaca}
\affiliation{The Observatories of the Carnegie Institution for Science, 813 Santa Barbara Street, Pasadena, CA 91101, USA}

\author[0000-0002-0572-8012]{Vedant Chandra}
\affiliation{Center for Astrophysics $|$ Harvard \& Smithsonian, 60 Garden Street, Cambridge, MA 02138, USA}

\author[0000-0002-6800-5778]{Jiwon Jesse Han}
\affiliation{Center for Astrophysics $|$ Harvard \& Smithsonian, 60 Garden Street, Cambridge, MA 02138, USA}

\author[0000-0002-9280-7594]{Benjamin D. Johnson}
\affiliation{Center for Astrophysics $|$ Harvard \& Smithsonian, 60 Garden Street, Cambridge, MA 02138, USA}

\author[0000-0003-3997-5705]{Rohan~P.~Naidu}
\altaffiliation{NASA Hubble Fellow}
\affiliation{MIT Kavli Institute for Astrophysics and Space Research, 77 Massachusetts Ave., Cambridge, MA 02139, USA}

\author[0000-0001-5082-9536]{Yuan-Sen Ting (丁源森)}
\affiliation{Department of Astronomy, The Ohio State University, Columbus, OH 43210, USA}
\affiliation{Center for Cosmology and AstroParticle Physics (CCAPP), The Ohio State University, Columbus, OH 43210, USA}
\affiliation{Research School of Astronomy \& Astrophysics, Australian National University, Cotter Rd., Weston, ACT 2611, Australia}

\begin{abstract}
\noindent Our understanding of the assembly timeline of the Milky Way has been transforming along with the dramatic increase in astrometric and spectroscopic data available over the past several years.  Many substructures in chemo-dynamical space have been discovered and identified as the remnants of various galactic mergers.  To investigate the timeline of these mergers we select main sequence turn off \& subgiant stars (MSTOs) from the H3 survey, finding members in seven metal poor components of the halo: GSE, the Helmi Streams, Thamnos, Sequoia, Wukong/LMS-1, Arjuna, and I'itoi.  We also select out the metal poor in situ disk to facilitate comparison to the evolution of the Milky Way itself at these early epochs.  We fit individual isochrone ages to the MSTOs in each of these substructures and use the resulting age distributions to infer simple star formation histories.  For GSE we resolve an extended star formation history that truncates $\approx10$ Gyr ago, as well as a clear age---metallicity relation.  From this age distribution and measured star formation history we infer that GSE merged with the Milky Way at a time $9.5-10.2$ Gyr ago, in agreement with previous estimates.  We infer that the other mergers occurred at various times ranging from $9-13$ Gyr ago, and that the metal poor component of the disk built up within only a few billion years.  These results reinforce the emerging picture that both the disk and halo of the Milky Way experienced a rapid assembly.
\end{abstract}

\section{Introduction}
The assembly history of the Milky Way is quickly being unveiled.  Precision proper motions and parallaxes from \emph{Gaia} DR2 \& DR3 \citep{Gaia_2018a, Gaia_2021} combined with radial velocities and elemental abundances from large scale spectroscopic surveys \citep[e.g.,][]{Steinmetz_2006, Yanny_2009, Gilmore_2012, Deng_2012, de_Silva_2015, Majewski_2017, Conroy_2019, Recio_Blanco_2023, De_Angeli_2023} have provided an unprecedented view into our Galaxy's dynamical and chemical evolution.  These data have enabled the discovery of dozens of substructures---coherent groups of stars in chemo-dynamical space---many of which are expected to be the remnants of satellite galaxies that have merged with the Galaxy.  Stars in a satellite share a star formation and chemical enrichment history and are deposited on similar orbits as the satellite is tidally disrupted by the Milky Way, creating one or potentially multiple new substructures \citep{Johnston_1996, Bullock_2005}.

Arguably the best example of this is the Sagittarius dwarf galaxy, which is in the process of being disrupted and still has an intact core along with a stream of tidally stripped stars that encircles the Milky Way multiple times \citep{Ibata_1994, Majewski_2003, Belokurov_2006}.  Sagittarius' tidal disruption began recently and is still ongoing \citep{Laporte_2018} and therefore its debris remains spatially coherent.  The remnants of mergers that occurred at earlier times will instead be phase mixed, no longer forming coherent structures in position space alone \citep{Johnston_1996, Helmi_1999a}.  

Along with Sagittarius, another of the earliest discovered merger remnants are the Helmi Streams (HS) \citep{Helmi_1999b} which were first identified by the coherent velocity structure of their stars in the solar neighborhood.  More recent analysis of the HS \citep{Koppelman_2019b} found their progenitor was a relatively massive satellite that merged before Sagittarius did, and whose stars now contribute $\gtrsim10\%$ of the halo population.  There have been many more substructures identified in the past few years that are believed to have come from smaller and less massive mergers, each contributing at most only a few percent to the field population of halo stars \citep{Myeong_2019, Koppelman_2019a, Naidu_2020, Yuan_2020b}.  With the exception of Sagittarius \citep{Schonrich_2018, Laporte_2018, Antoja_2018, Ruiz_Lara_2020}, all of these minor mergers had such unequal mass ratios that they are not expected to have had a significant impact on the wider Milky Way's evolution or morphology.

The remnant of a major merger has also been identified, the \emph{Gaia}-Sausage/Enceladus (GSE) merger whose debris was dramatically revealed after the release of \emph{Gaia} DR2 \citep{Belokurov_2018, Helmi_2018}.  GSE is now understood to be the most significant merger in the Milky Way's history.  It was low total mass ratio ($\sim$1:3) and contributed up to roughly 1/2 of all halo stars \citep{Feuillet_2020, Naidu_2020, Naidu_2021} and around 20\% of the Milky Way's entire globular cluster system \citep{Massari_2019}.  

In addition to being the dominant contributor to the stellar halo, the GSE merger was massive enough to have dramatically altered the star formation history and morphology of the Milky Way's disk \citep{Walker_1996, Han_2023a}.  A variety of possible impacts on the Milky Way have been attributed to the GSE merger: the dynamical heating of Galaxy's original disk to form the modern thick disk \citep{Wyse_2001}, scattering orbits further to create the in situ halo/Splash population (\citealt{Bonaca_2017, Belokurov_2020, Bonaca_2020}, but see also \citealt{Amarante_2020, Beraldo_e_Silva_2020}), and inducing the disk's abundance bimodality via gas infall and a metallicity reset \citep{Palla_2020, Ciuca_2023}.

Properly establishing causality between the GSE merger and these transformations to the Milky Way requires precise measurements of the timing of these events.  Ideally this would be done by independently measuring the star formation histories (SFHs) of GSE and the Milky Way and then connecting changes in those SFHs to the aforementioned chemical and morphological transformations.  This in turn requires estimates of the ages of their stars.  There is a lack of age information compared to chemical or dynamical information for halo stars, because age estimation is only viable for a subset of evolutionary phases and because age estimation is very sensitive to systematics and requires relatively higher signal to noise data.  Studies that leverage age information thus far have therefore been limited in one way or another.  Fitting the color magnitude diagram (CMD) of an entire halo population requires precise individual parallaxes, limiting the usable volume.  This works for GSE due to it being the dominant halo component local to the Sun \citep{Gallart_2019}, but the limited volume prevents the study of the other smaller merger remnants that are not as well represented close to the Sun.  Asteroseismology of giant stars has the promise of delivering precise ages out to much larger distances, but the designs and sensitivity of existing time domain surveys have limited seismic studies of the halo to at most a handful of stars \citep{Grunblatt_2021, Montalban_2021, Borre_2022}.  Fitting individual stellar isochrones to spectroscopic survey data fills the gap between these two methods, being effective out to larger distances than entire CMD fitting while still having enough stars with which to build SFHs for the smaller halo populations.  

In this work we measure isochrone ages for main sequence turn off \& subgiant branch (hereafter simply MSTO) members of the metal poor halo substructures observed by the H3 survey, and using those ages we construct star formation histories (SFHs) for the progenitors of each substructure, with a particular focus on GSE.  In Section \ref{section:data} we detail the survey observations and data set used, in Section \ref{section:methods} we explain in detail the method of fitting isochrones (Section \ref{subsec:isochrone ages}) and star formation histories (Section \ref{subsec:sfh fitting}).  Section \ref{section:results} gives an overview of the inferred age distribution and star formation history for each substructure, and in Section \ref{section:discussion} we address what these results imply about GSE's impact on the Milky Way and on the Galaxy's overall assembly history.  Lastly we summarize our results in Section \ref{section:conclusion}.  Appendices \ref{app:isochrone mocks} \& \ref{app:sfh fitting} provide additional details and supplementary information about the fitting procedures in Sections \ref{subsec:isochrone ages} \& \ref{subsec:sfh fitting}.

\section{Data}
\label{section:data}

\subsection{H3 Survey}
\label{subsec:h3}
\begin{figure}
\epsscale{1.0}
\plotone{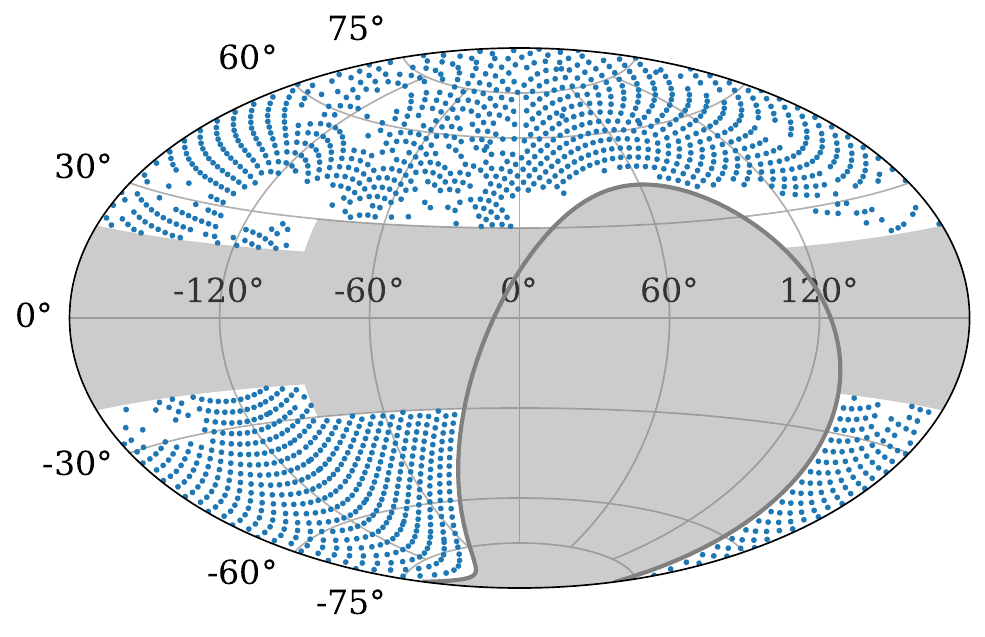}
\caption{Current footprint achieved by the H3 survey, in Galactic coordinates.  Target tiles are restricted to $|b|>30^\circ$ towards the galactic center and $|b|>20^\circ$ towards the anti-center by survey design. The survey is conducted from the MMT observatory on Mt. Hopkins, Arizona, USA so there is also a geographic constraint of $\delta>-20^\circ$, indicated by the bold line.\label{fig:h3 footprint}}
\end{figure}

We use data from the H3 survey \citep{Conroy_2019}, a high resolution spectroscopic survey optimized for the study of the stellar halo.  The survey is conducted using the Hectochelle fiber-fed multi-object spectrograph on the MMT, taking spectra around the Magnesium triplet $(5150\angstrom-5300\angstrom)$ at a resolution of $R\sim32,000$ \citep{Fabricant_2005, Szentgyorgyi_2011}.  H3 tiles the sky above the galactic plane ($|b| > 20^\circ$ towards the Galactic anticenter, $|b| > 30^\circ$ towards the Galactic center, $\delta > -20^\circ$), targeting stars in the magnitude range of Pan-STARRS $15 < r < 18$, with \emph{Gaia} DR2 or eDR3 parallax $\varpi < 0.3$.  The survey footprint in plotted in Figure \ref{fig:h3 footprint}.  The survey first began in the fall of 2017 and at the time of this analysis has observed a total of $323,000$ stars.

The survey's main parallax and magnitude criteria are designed to preferentially target intrinsically luminous giant stars at large distances.  H3's targeting scheme also includes additional filler targets at higher parallax to maximize fiber usage during each pointing.  In this work we take advantage of these filler targets, measuring isochrone ages for the MSTO stars found in halo substructures and using them to infer the star formation histories of the Galaxy's metal poor components.  

We select our sample from the H3 v4.1 catalog, requiring spectroscopic S/N $>7$ and no flags indicating a bad quality fit.  The v4.1 catalog adopts a flat prior on stellar age and distance during the fitting process (see Section \ref{subsec:isochrone ages}),  in contrast to the fiducial v4.0 catalog which adopts a multi-component prior on both age and distance that relies on a Galactic density model.  From this initial sample we select out MSTO stars using a cut of $3.5<$ \logg\ $<4.3$ (see Appendix \ref{app:isochrone mocks} for justification of this selection).  We then refit the stars in this MSTO subsample using a wider uniform age prior that extends from 4 to 20 Gyr (see Section \ref{subsec:isochrone ages}).  Lastly, we impose additional quality cuts on the MSTO subsample to ensure the age estimates are precise and reliable.  For the MSTOs we additionally require that $\varpi/\sigma_\varpi > 5$, the reduced $\chi^2_\mathrm{spec} < 2.5$, the reduced $\chi^2_\mathrm{phot} < 2.5$, \emph{Gaia} RUWE $<$ 1.3, and the rotational velocity $V_\mathrm{rot} < 3$ km s$^{-1}$.

\subsection{Substructure Selection}
\label{subsec:substructure selection}
\begin{figure*}[ht!]
\plotone{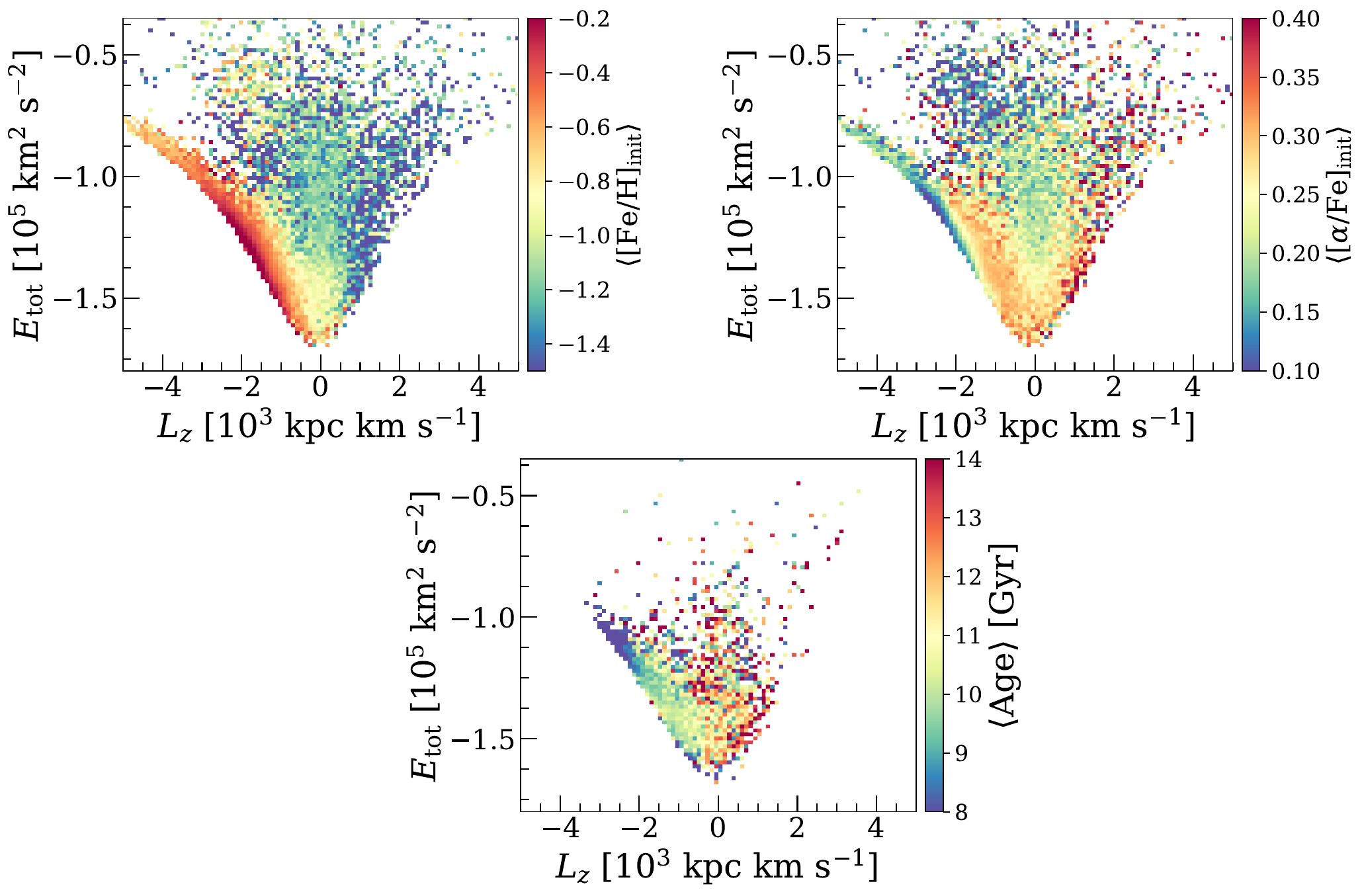}
\caption{Distribution of all H3 stars with spectral S/N $>7$ in the $E-L_z$ plane, colored by the mean \fehi\ (\emph{Left}) and mean \afei\ (\emph{Right}) in each bin.  Prograde orbits in this coordinate convention have negative $L_z$.  It is clear that there is a high degree of substructure; debris from GSE dominates the radial halo at moderate energies, but a variety of other structures are visible flanking GSE, contrasting in their orbits and abundances.   \emph{Bottom}: Distribution of H3 MSTO stars with spectral S/N $>7$ in the $E-L_z$ plane, colored by the mean Age in each bin.\label{fig:E Lz substructure}}
\end{figure*}

There are known to be $\approx15-20$ distinct substructures within the Milky Way's stellar halo, with previous studies identifying $\approx$ 13 structures represented within the H3 survey's data \citep{Johnson_2020, Bonaca_2020, Naidu_2020, Naidu_2022b}.  Figure \ref{fig:E Lz substructure} shows the Galaxy in chemo-dynamical space as seen by H3, with $E-L_z$ space binned into cells that are colored by the mean metallicity, alpha element enhancement, or age of the stars in that cell.  These projections of chemo-dynamical space clearly demonstrate the rich substructure present in the Galaxy's stellar halo.  In this work we focus on the metal poor structures that represent the debris of now disrupted dwarf galaxies.  We disregard three kinematically cold streams known to be observed by H3 (Sagittarius, Cetus, and Orphan/Chenab).  Though these substructures are themselves associated with disrupted dwarfs, each is sufficiently distant such that they are only represented among H3's giants and thus are naturally excluded when we restrict our sample to only main sequence turn off stars.  

We also elect to disregard all metal rich structures, essentially all of which are associated with in situ Galactic disk populations.  These structures include the high$-\alpha$ and low$-\alpha$ disks, the in situ halo \citep{Bonaca_2017, Bonaca_2020, Belokurov_2020}, and Aleph \citep{Naidu_2020, Horta_2022, Han_2023a}.  The stellar ages and star formation histories of these populations (with the exception of Aleph) have already been well studied, so we instead focus on the metal poor structures whose stellar age information is more novel.  We exclude the metal rich structures using a cut in \fehi$-$\afei\ abundance space, similar to the one used in \citet{Naidu_2020} but modified slightly to increase the purity of the metal poor sample at low \afei.  The subscript denotes that we use a star's \emph{initial} abundances, as opposed to it's present day surface abundances which may have changed over it's lifetime (see Section \ref{subsec:isochrone ages}).

\begin{figure*}
\plotone{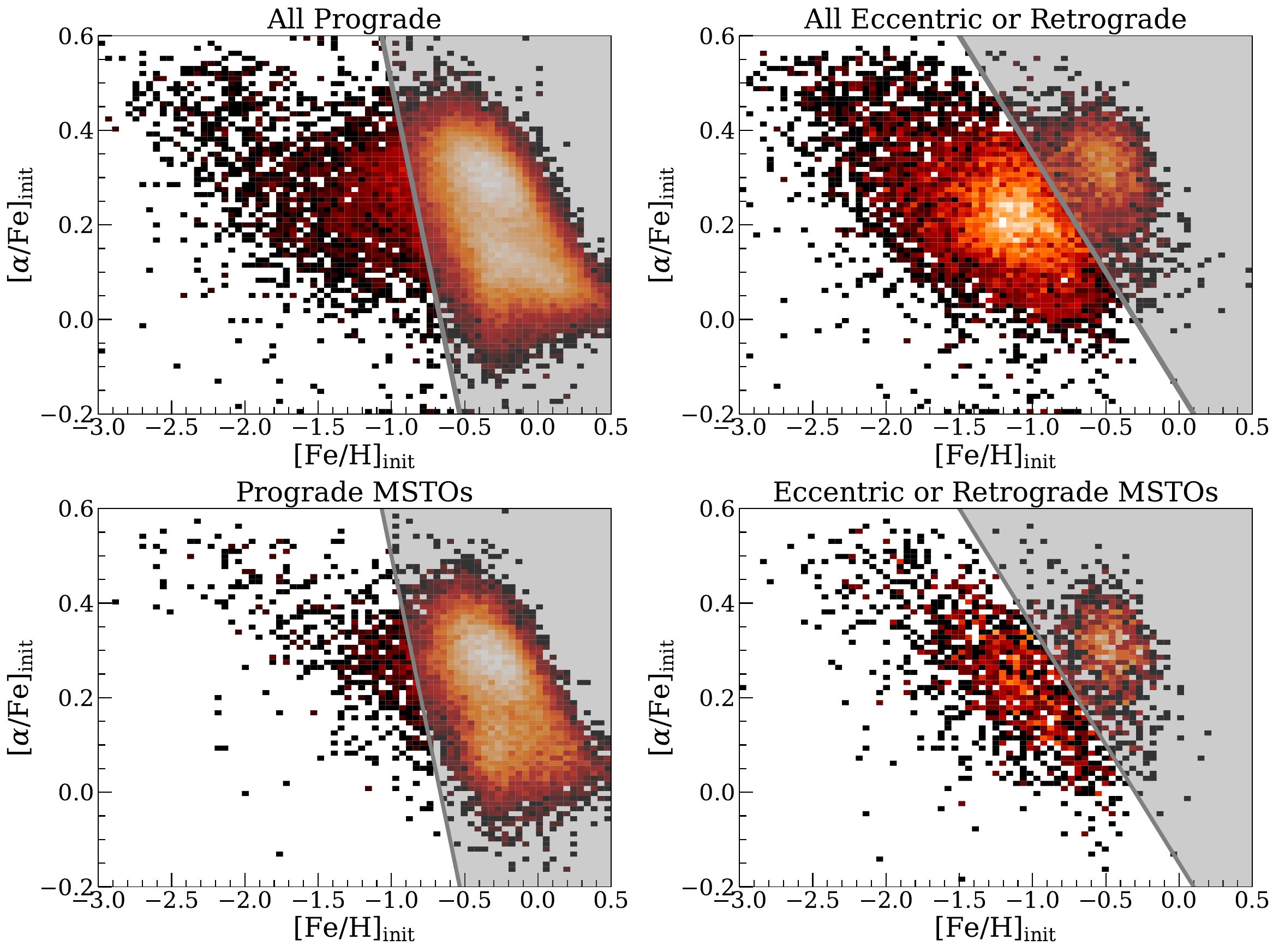}
\caption{\fehi\ vs \afei\ abundance diagram demonstrating the abundance cut used to remove stars associated with metal rich in situ populations.  All stars found in the upper right shaded region are excluded from all subsequent analyses.  \emph{Top row}: All H3 stars in with S/N $>7$.  \emph{Bottom row}: Only the primary MSTO sample ($3.5<$ \logg\ $<4.3$).  \emph{Left column}: Prograde stars with $e < 0.7$.  \emph{Right column}: Eccentric stars defined as having $e>0.7$ and also retrograde stars with $e<0.7$.  The definition of the metal rich region varies slightly between the left and right columns.\label{fig:metal rich exclusion}}

\end{figure*}

Figure \ref{fig:metal rich exclusion} show H3 stars in the \fehi$-$\afei\ plane.  The top row includes all H3 stars with spectral S/N $>7$, while the bottom row includes only the MSTOs that satisfy all of the additional quality cuts.  The metal rich region excluded by the abundance cut is highlighted in gray.  One version of the cut is applied to stars with prograde and low eccentricity orbits (the left column), while the other version of the cut is applied to stars with either highly eccentric or retrograde orbits (the right column).  Note that prograde orbits in this coordinate convention have $L_z < 0$.

\begin{itemize}
\item[i)] $[\alpha/\mathrm{Fe}]_\mathrm{init} > 0.2 - 1.5([\mathrm{Fe/H}]_\mathrm{init} + 0.8)$
\item[ii)] $(e< 0.7)~\&~(L_z<0)$
\item[] OR
\item[i)] $[\alpha/\mathrm{Fe}]_\mathrm{init} > 0.2 - 0.5([\mathrm{Fe/H}]_\mathrm{init} + 0.7)$
\item[ii)] $(e > 0.7)$
\item[] OR
\item[i)] $[\alpha/\mathrm{Fe}]_\mathrm{init} > 0.2 - 0.5([\mathrm{Fe/H}]_\mathrm{init} + 0.7)$
\item[ii)]  $(e< 0.7)~\&~(L_z>0)$
\end{itemize}

\begin{figure}
\epsscale{1.0}
\plotone{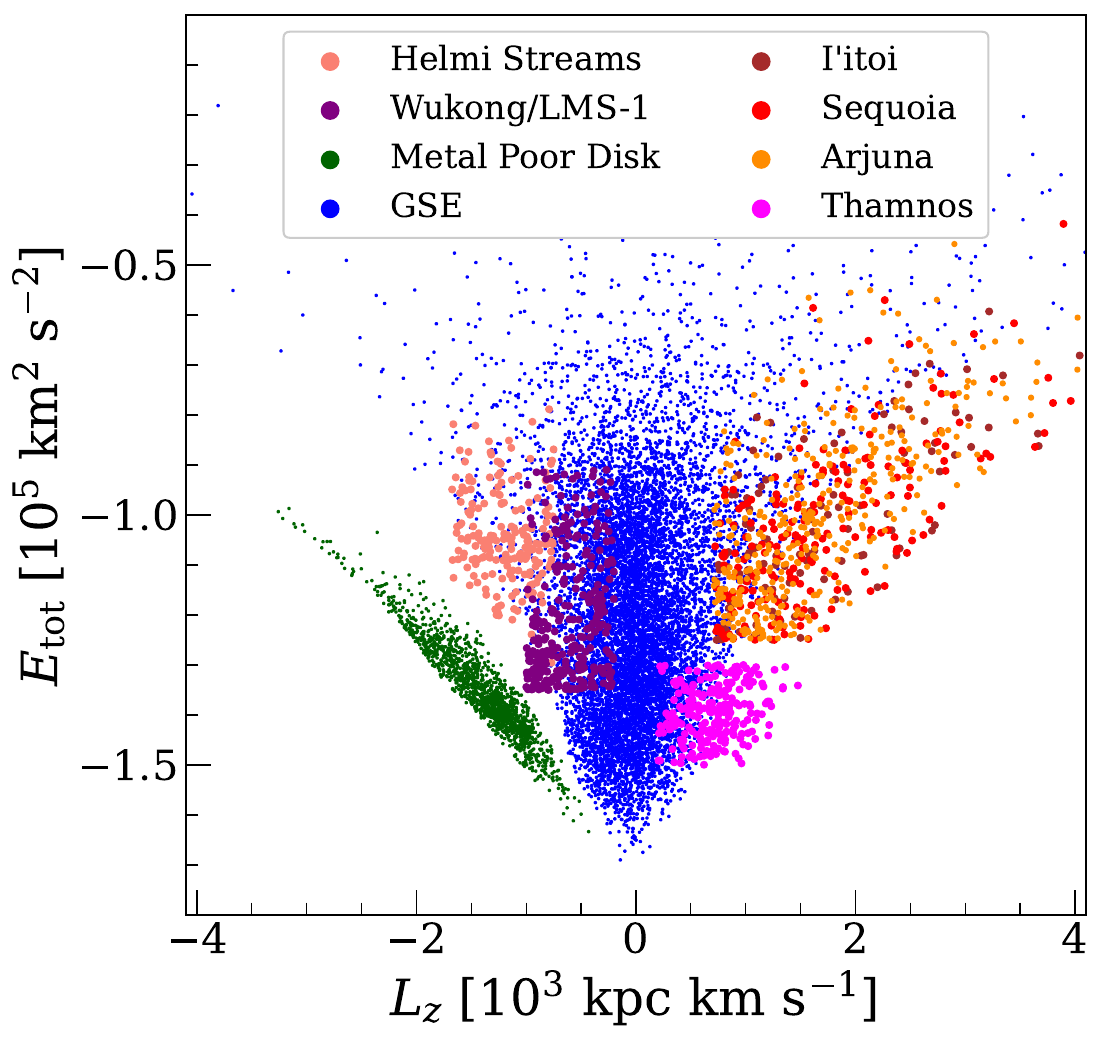}
\caption{$E-L_z$ plot of stars with S/N $>7$ in each of the eight substructures we consider in our analysis.  This figure includes stars of all evolutionary phases and is not restricted to only MSTOs.  Each substructure with the exception of the metal poor disk (dark green) is associated with a disrupted dwarf galaxy that merged with the Milky Way in the past.  This and subsequent figures share a color coding for each substructure.\label{fig:substructure schematic}}
\end{figure}

Adopting the rest of the selection cuts from \citet{Naidu_2020} leaves us with eight metal poor substructures: GSE, Thamnos, Arjuna, Sequoia, I'itoi, Wukong/LMS-1, the Helmi Streams, and a metal poor component of the Galactic disk.  Figure \ref{fig:substructure schematic} shows the distribution of each of these structure in the $E-L_z$ plane, using all H3 stars with S/N $>7$ and not only the MSTOs.  We note that \citet{Naidu_2021} argues that Arjuna is not a distinct dwarf galaxy but is instead the high energy retrograde debris of GSE, but we choose to consider Arjuna on its own when performing our analysis, and will return to the question of its origin in Section \ref{section:discussion}.

\section{Methods}
\label{section:methods}

\subsection{Isochrone Ages}
\label{subsec:isochrone ages}

\begin{figure*}
\epsscale{1.0}
\plotone{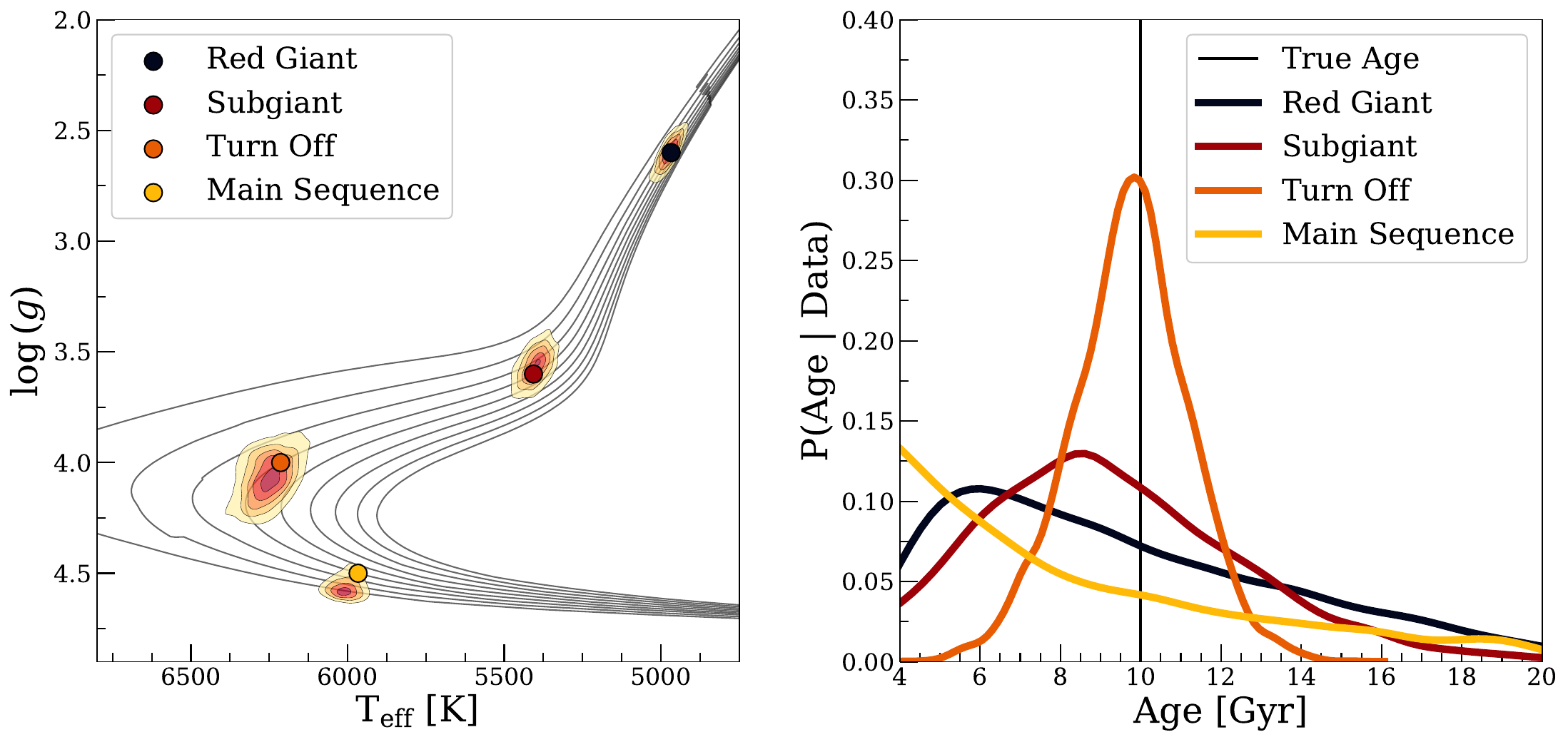}
\caption{\emph{Left:} \texttt{MINESweeper} \Teff$-$\logg\ posteriors for mock observations with an age of 10 Gyr at different evolutionary phases.  In the background we plot isochrones from 4 to 20 Gyr in steps of 2 Gyr.  \emph{Right}: The marginalized age posteriors for those different evolutionary phases.  Given the typical S/N of H3 data, the main sequence and red giant branch are not useful for isochrone age dating.  Only the main sequence turn off and subgiant branch have isochrone ages that are both sufficiently precise and unbiased for our purposes.\label{fig:kielposteriors}}
\end{figure*}

The H3 spectra are analyzed using \texttt{MINESweeper}, a Bayesian inference stellar fitting code \citep{Cargile_2020}.  \texttt{MINESweeper} infers stellar parameters by simultaneously fitting a star's spectrum, broadband photometry, and \emph{Gaia} parallax ($\varpi$) against a library of isochrone and spectral models.  In each likelihood call an evolutionary state (EEP \citep[see][]{Dotter_2016}, initial mass, \fehi, \afei) is sampled from the isochrone grid, then the predicted atmospheric parameters of this state (\Teff, \logg, \logR, \feh, \afe) are combined with a sampled distance/parallax and extinction coefficient in the $V$ band (\Av) to predict the star's spectral energy distribution (SED) and compare it to the data using the framework of \texttt{ThePayne} \citep{Ting_2019}.  Each star's data consists of its Hectochelle spectrum ($\lambda\sim5150-5300$ \AA), and the following photometry: \emph{Gaia} G, BP, RP, Pan-STARRS \emph{grizy} , 2MASS \emph{J, H, $K_s$}, UNWISE W1 \& W2, and SDSS \emph{ugriz}.  The \emph{Gaia} and Pan-STARRS photometry is available for all of our sample as part of the H3 survey's input catalog, while 2MASS, UNWISE, and SDSS photometry is available for $>99\%, >96\%$, and $>81\%$ of our MSTO sample respectively.

We adopt the \texttt{MIST} v2.0 isochrone grid (\citet{Choi_2016}, Dotter et al. in prep.), together with the most up to date C3K spectral grid with a custom tuned line list \citep[see][]{Ting_2019}.  These two grids are constructed self-consistently with matching physical inputs e.g.,  both adopt the \citet{Asplund_2009} solar abundance pattern.  This avoids potential systematics that can be introduced when combining mismatched isochrones and spectral grids.  The \texttt{MIST} v2.0 models include a prescription for the diffusion of heavy elements out of the outer layers of a star, introducing a distinction between a star's initial abundances (\fehi, \afei) and the surface abundance (\feh, \afe), the latter of which is the abundance used when predicting the spectrum and photometry.  The distinction between initial and surface abundances becomes particularly significant around the main sequence turn off and subgiant branch (with offsets up to $\Delta$[Fe/H]$=0.25$ for old metal poor stars), and therefore diffusion cannot be neglected when using these phases to infer isochrone ages.  Ignoring the effects of diffusion will introduce 10---20\% error in the inferred ages of stars, enough to dominate over typical statistical uncertainties \citep[see Appendix \ref{app:isochrone mocks};][]{Dotter_2017}.  Since it is the initial abundances that are physically relevant when interpreting star formation histories and chemical evolution, we focus on initial abundances throughout the rest of this paper, denoted as \fehi\ and \afei. 

\texttt{MINESweeper} allows for a variety of priors to be placed on any of the directly sampled or inferred parameters.  As mentioned above the fiducial H3 pipeline utilizes a complex prior on age and distance based on a three component model of the galactic disk and halo \citep{Conroy_2019, Cargile_2020}.  This prior aids in distance inference for the distant giants that form the core of the H3 sample, but because stellar ages are our primary concern in this work we instead adopt a uniform prior on both age and distance.  We choose a uniform prior on age that extends from 4 to 20 Gyr, allowing for ages older than the accepted age of the universe \citep{Planck_2020}.  This is done to avoid an artificial pile up of stars at the old edge of the prior; the median statistical age uncertainty of our MSTO sample is $\approx11\%$ and so age measurements of genuinely old stars are expected to scatter up to a few Gyr.  We therefore must allow for ages several Gyr past the age of the universe in order to sufficiently resolve star formation histories at those early times.  The tail of ages past 14 Gyr is a product of both the expected statistical scatter combined with a variety of potential outlier sources.  See Section \ref{subsec:sfh fitting} for details on how we treat these anomalously old stars when measuring star formation histories, and see Appendix \ref{app:isochrone mocks} for an example of one source of outliers: unresolved binary companions.

We adopt a Kroupa IMF prior on the initial mass \citep{Kroupa_2001} and uniform priors on equal evolutionary point \citep[EEP; see][]{Dotter_2016}, \fehi, and \afei.  The extinction coefficient \Av\ is given a Gaussian prior centered on the value measured in the \citet{Schlegel_1998} dust map, assuming $R_V=3.1$ and applying the 14\% downward correction to the normalization recommended by \citet{Schlafly_2011}.  The priors on the rest of the parameters are left unchanged from the defaults used in H3.  \citet{Cargile_2020} goes into further detail about the fitting procedure and the rest of these adopted priors.  

Figure \ref{fig:kielposteriors} demonstrates the results of applying our isochrone fitting procedure to mock observations.  On the left we plot the fitted \Teff$-$\logg\ posterior for mock stars with the same age but different evolutionary phases, and on the right we plot the corresponding marginalized age posteriors.  When we adopt spectral and parallax S/N typical of H3 stars the age inference on the main sequence and red giant branch is very imprecise, with the uncertainty on the fitted ages reaching upwards of 50\%.  In these cases the ages are not only imprecise but are also \emph{biased}, with the age posterior becoming dramatically offset from the true age.  We are therefore restricted to only use stars around the main sequence turn off and subgiant branch evolutionary phases, where the fitted age is not significantly biased from the true age.  Appendix \ref{app:isochrone mocks} explores this age recovery bias in more depth and provides further justification for the $3.5<$ \logg\ $<4.3$ cut we use to choose stars with reliable ages.

\subsection{Selection Function}
\label{subsec:selection function}

\begin{figure}
\epsscale{1.0}
\plotone{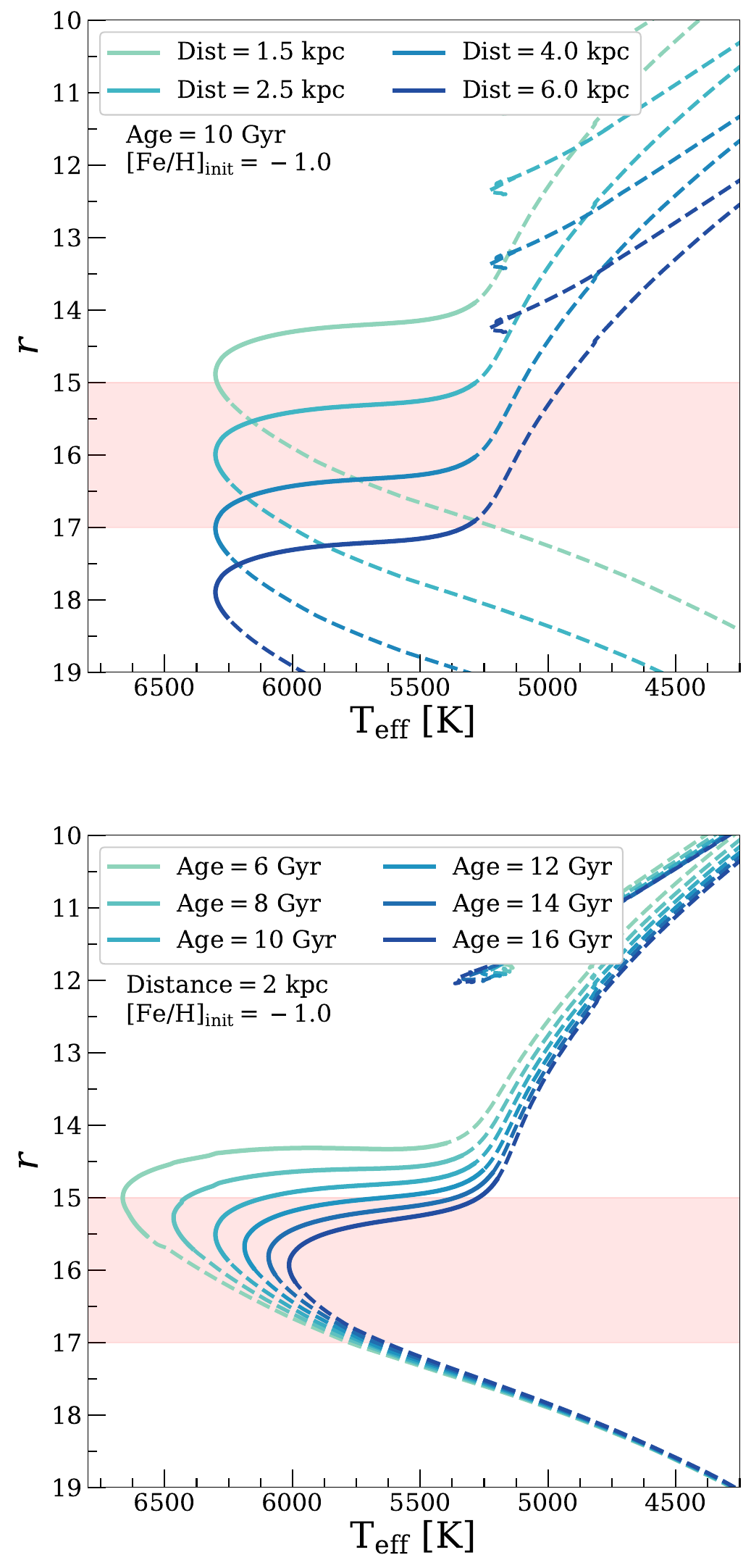}
\caption{Demonstration of the selection function applied to theoretical isochrones.  The portions of an isochrone that satisfy $3.5 <$ \logg\ $< 4.3$ are plotted as solid lines while the rest of the isochrone is plotted as a dashed line.  The effective magnitude window ($15 < r < 17$) is plotted as a red shaded region.  \emph{Top}:  For a given isochrone (e.g., a typical halo population with Age $=10$ Gyr, \fehi $= -1.0$, \afei$=+0.2$, \Av$=0.1$) there is a specific range of distances at which MSTOs will satisfy our magnitude selection.  This range of distances creates an effective survey volume that is incorporated into the selection function.  \emph{Bottom}:  The same isochrones but now fixed at a distance of 2 kpc and varying in age.  We can see that the luminosity of MSTOs vary with age and therefore the effective survey volume will also depend on age.\label{fig:selection function isochrones}}
\end{figure}

The selection function of our MSTO sample can be understood as the combination of three primary components.  First is the magnitude selection of the H3 survey, which is nominally $15 < r < 18$.  However, our requirement for spectral S/N $> 7$ ends up changing the limiting faint magnitude, making it a function of the star's targeting tile and date of observation.  The survey adopts a uniform 30 minute exposure time for each field, so as observing conditions vary from night to night and tile to tile (e.g., airmass, the Moon's brightness) the S/N achieved in that time for a fixed apparent magnitude will change.  For each tile and date we calculate the effective limiting magnitude implied by our S/N requirement, but generally S/N $> 7$ implies $r\approx17$ as a limiting faint magnitude.  In addition to the magnitude selection we also apply a temperature cut of $\mathrm{T}_\mathrm{eff}<7000$ K because the spectral and isochrone fits begin to lose fidelity at higher temperatures, and lastly we apply a cut of $3.5<$ \logg $<4.3$ to select out MSTO stars with precise and reliable ages.  

Figure \ref{fig:selection function isochrones} shows how these selections apply to model isochrones in \Teff$-r$ magnitude space.  In the top panel we plot the isochrone of a typical halo population at various distances, demonstrating how our magnitude criterion creates an effective survey volume that spans distances from $1.5-6$ kpc.  Outside of this volume MSTOs are either too bright or too faint to be targeted by H3.  Stars that are within but near the edges of this volume must be weighted more heavily because our magnitude criterion begins to exclude parts of our \logg\ selection.  In the bottom panel we plot isochrones at various ages while keeping the distance fixed.  The luminosity of our \logg\ selection decreases with age and therefore the edges of the effective survey volume will depend on the age of the population we are trying to observe.  The initial metallicity \fehi\ also has a significant effect on the luminosity and so the general trend is that stars that are older or more metal rich must be weighted more heavily because these stars are intrinsically fainter and therefore will only be selected from a smaller effective volume. 

Applying these three cuts to an isochrone of a given Age, \fehi, \afei, Distance, and $A_V$ selects out just the portion of the isochrone around the main sequence turn off and subgiant branch, provided they are within the permitted magnitude and temperature ranges.  Integrating along this selected portion using a Kroupa IMF \citep{Kroupa_2001} then gives us what fraction of a simple stellar population we expect to satisfy our selection cuts and therefore be observed.  This can be understood using Equation \ref{eq:f_IMF}: 
\begin{equation}
\label{eq:f_IMF}
f^N_\mathrm{IMF} = \int_{M_l}^{M_u} \phi dM~~~\mathrm{OR}~~~f^M_\mathrm{IMF}=\int_{M_l}^{M_u} M\phi dM,
\end{equation}
where $\phi$ is the IMF, and $M_l$ \& $M_u$ are the lower and upper mass limits that satisfy our magnitude, temperature, and \logg\ cuts.  $M_l$ \& $M_u$ generally are functions of the Age, \fehi, \afei, Distance, and \Av.  The relevant weight to account for the effects of the selection function is then simply $W=(f_\mathrm{IMF})^{-1}$.  When counting the number of stars (e.g., Figures \ref{fig:n age hists} \& \ref{fig:n amrs}) we weight by $W=(f^N_\mathrm{IMF})^{-1}$, and when measuring a mass (e.g., star formation histories) we weight by $W=(f^M_\mathrm{IMF})^{-1}$.  

\subsection{SFH Fitting}
\label{subsec:sfh fitting}
Using the isochrone ages for each substructure's MSTO stars, we fit simple star formation histories (SFHs) to the resulting age distributions.  The age posteriors from our individual \texttt{MINESweeper} fits are well behaved and largely symmetric, so we approximate each star's age posterior as a truncated Gaussian distribution bounded between 4 and 20 Gyr, parameterized by the mean and standard deviation of the sampled age posterior.  Given our typical age uncertainties and small number of MSTOs per substructure, we do not have sufficient information to discern between different functional shapes for the SFH in a meaningful way.  We therefore make a simplifying assumption about the shape and treat each substructure's SFH as a Gaussian distribution truncated between 4 and 14 Gyr.

Letting $\psi(t | \theta)$ denote the star formation rate as a function of time and $\theta$ denote the set of parameters describing $\psi$, a star's likelihood contribution is given by Equation \ref{eq:individual likelihood}:
\begin{equation}
\label{eq:individual likelihood}
\mathcal{L}_i(\theta) = W_i \times \int_{4~\mathrm{Gyr}}^{20~\mathrm{Gyr}} \psi(t|\theta) \tau(t |\mu_i, \sigma_i) dt.
\end{equation}
The term $\tau(t|\mu_i, \sigma_i)$ in the integrand is the star's age posterior, and the entire integral is weighted by the selection function term $W_i$.  The total likelihood is then simply the product of all individual likelihoods $\mathcal{L}(\theta) = \prod_{i=1}^{N_\mathrm{substr}}\mathcal{L}_i(\theta)$.

The SFH function includes an outlier component in addition to primary component that we wish to infer.  This serves the purpose of guaranteeing all individual likelihoods to be non-zero, and it will also serve to absorb outliers like the anomalously old stars mentioned in Section \ref{subsec:isochrone ages}.  Again, Appendix \ref{app:isochrone mocks} offers a more in depth look at some potential sources of outliers, but the sources we consider are by no means exhaustive.  We instead choose to adopt a simple and uninformative outlier component, a uniform distribution spanning the entire range of ages from 4 to 20 Gyr.  This outlier component is added to the primary component that we are trying to infer with a membership fraction $f_\mathrm{out}$ that is included as a fitted parameter.  The expanded functional form of the SFH including the outlier term is given by Equation \ref{eq:expanded sfh}:
\begin{equation}
\label{eq:expanded sfh}
\begin{aligned}
\psi(t|\theta) &= (1-f_\mathrm{out})\psi_\mathrm{prim}(t|\theta) + f_\mathrm{out} \psi_\mathrm{out}(t) \\
\psi_\mathrm{prim}(t|\theta) &= \mathrm{Normal}(t|\mu_\mathrm{SFH}, \sigma_\mathrm{SFH},t_1=4, t_2=14)\\
\psi_\mathrm{out}(t) &= \mathrm{Uniform}(t|t_1=4, t_2=20).
\end{aligned}
\end{equation}
As mentioned above, we model the primary component of each substructure's SFH as a Gaussian distribution truncated between 4 and 14 Gyr, indicated by the bounds $t_1$ and $t_2$ in Equation \ref{eq:expanded sfh}.  These truncation bounds on $\psi_\mathrm{prim}(t|\theta)$ are where we chose to include a prior on the maximum age of stars given the age of the universe.

We perform each star formation fit with \texttt{emcee}, using 32 walkers running for 6000 steps and removing the first 1000 steps of burn in.  We adopt a uniform prior on the mean $\mu_\mathrm{SFH}$ from 4 to 14 Gyr.  For the membership fraction $f_\mathrm{out}$ we adopt a weak exponential prior that prefers higher membership fractions for $\psi_\mathrm{prim}$.  Lastly, for $\sigma_\mathrm{SFH}$ we adopt a uniform prior from 0.1 to 5 Gyr.  For two substructures (Arjuna and I'itoi) the fits are entirely uninformative on the parameter $\sigma_\mathrm{SFH}$, leaving its posterior prior dominated.  In these cases we redo the fit and instead adopt a scale invariant prior for $\sigma_\mathrm{SFH}$ ($\propto 1/\sigma_\mathrm{SFH}$) that ranges from 0.1 to 5 Gyr (see Appendix \ref{app:sfh fitting} and Figure \ref{fig:ArjunaIitoi_2priors}).

\section{Results}
\label{section:results}

\begin{figure*}
\epsscale{1.0}
\plotone{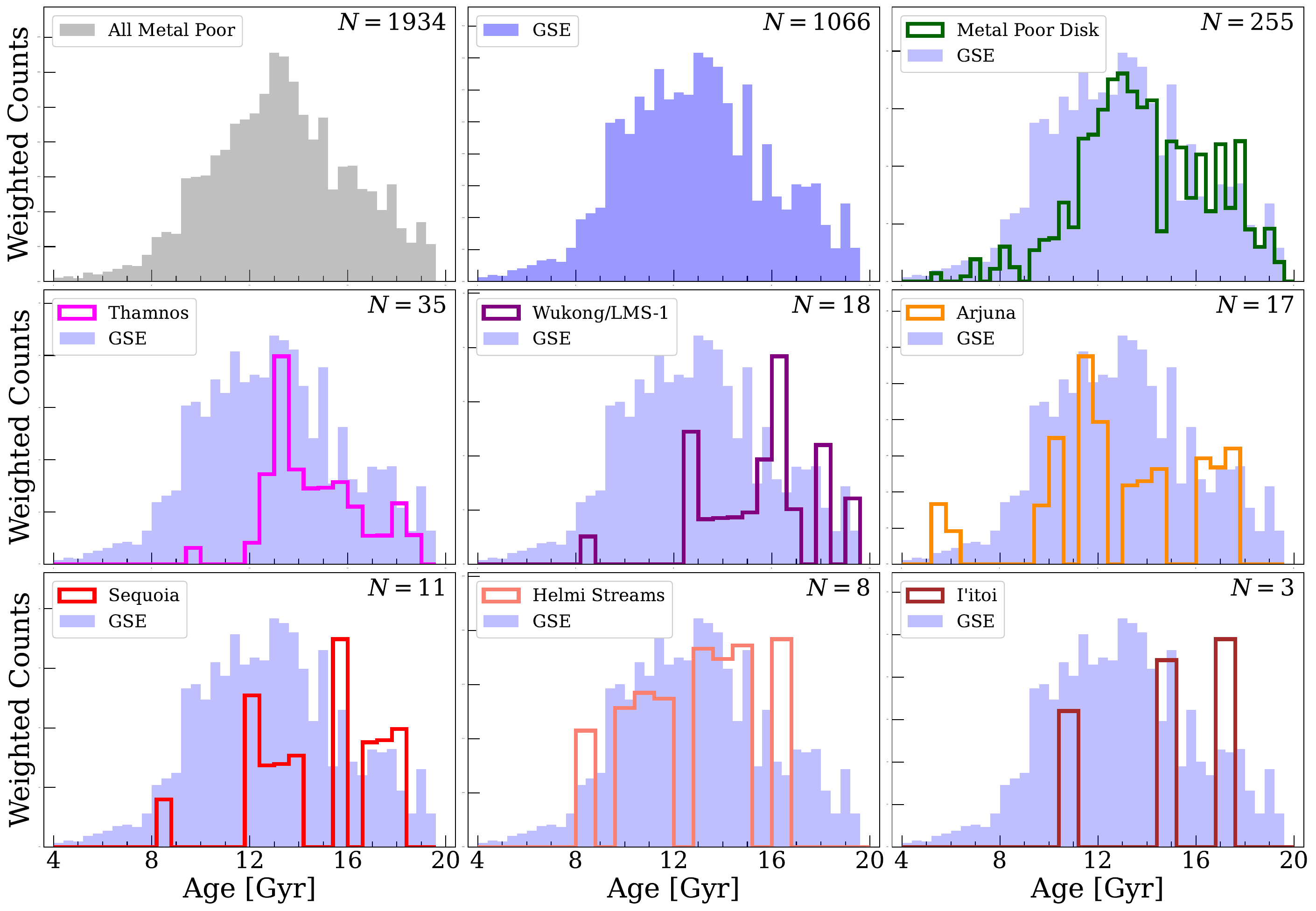}
\caption{Age distributions for the eight structures of interest, with counts normalized and weighted by the selection function.  Each panel has the age distribution of GSE plotted in the background for comparison, with an arbitrary normalization.  Wukong/LMS-1, the metal poor disk, Thamnos, Sequoia, and I'itoi are all on average older than GSE, indicating that their star formation likely truncated earlier.  The Helmi Streams exhibit a more extended SFH, and Arjuna appears to have a SFH only slightly older than that of GSE, with three young stars that could be outliers.\label{fig:n age hists}}
\end{figure*}

\begin{figure*}
\epsscale{1.0}
\plotone{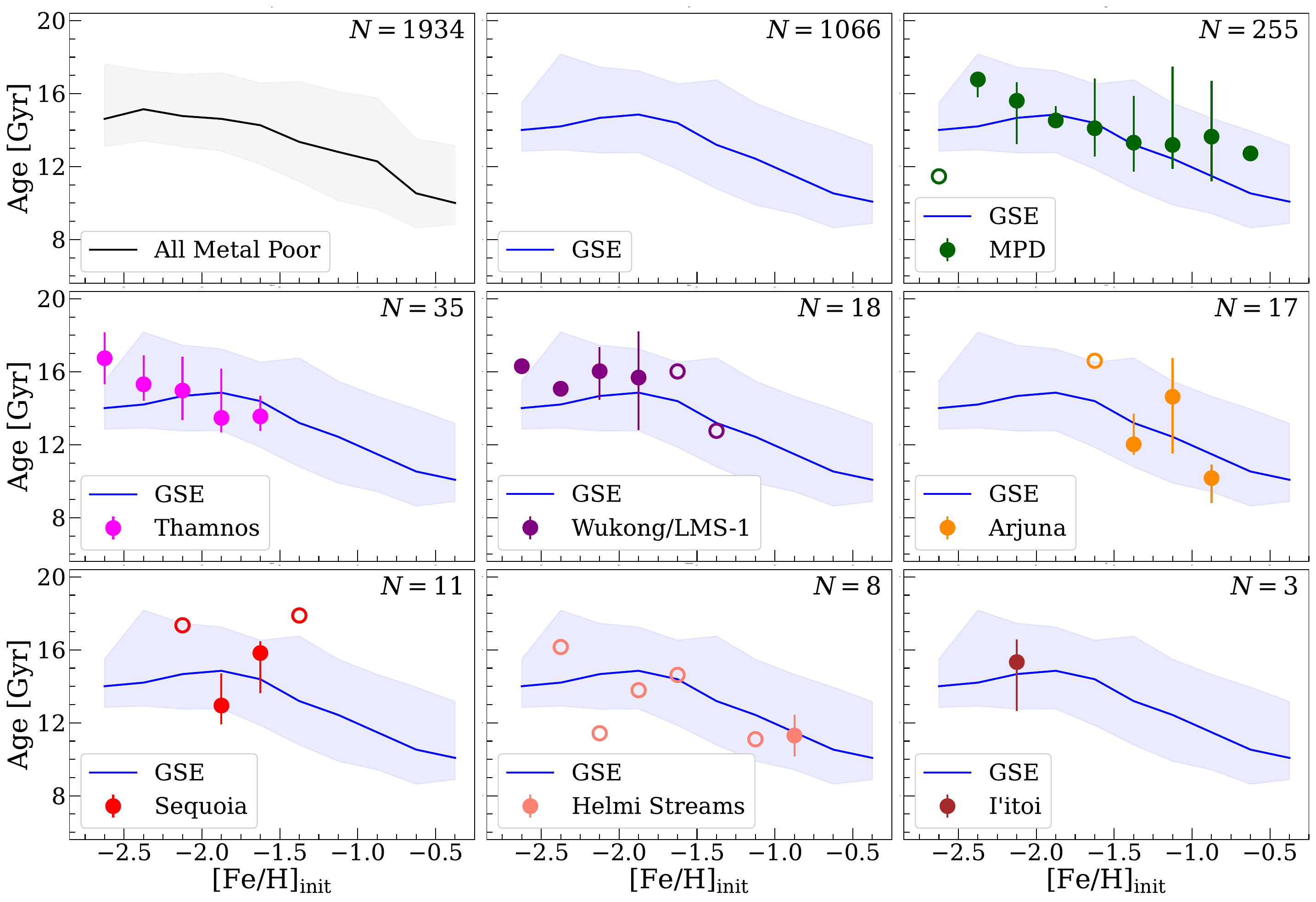}
\caption{Age$-$Metallicity relations for each substructure.  Just as in Figure \ref{fig:n age hists}, each panel has the AMR of GSE plotted in the background for comparison.  The data is binned in metallicity, with the filled points and error bars showing the weighted 16th, 50th, and 84th percentile age in that bin.  Metallicity bins with only a single star are plotted as open circles.  GSE exhibits a clean AMR, with it being largely flat until a metallicity of $\approx-1.7$, after which it then has a linear trend.  Most of the other substructures do not appear to have a significant AMR, with the exception of the MPD which at fixed metallicity tends to be older than GSE, and potentially the Helmi Streams which at fixed metallicity appears slightly younger.\label{fig:n amrs}}
\end{figure*}

\begin{deluxetable*}{l r l l l r}[t]
\tablecaption{Summary of the Substructure SFH Fits}
\tablehead{
\colhead{Substructure}    & \colhead{$N$} & \colhead{$\mu_\mathrm{SFH}$} & \colhead{$\sigma_\mathrm{SFH}$} & \colhead{$f_\mathrm{out}$} & \colhead{$\tau_{84}$ [Gyr]}}
\startdata
GSE             & 1066  & $12.71\pm^{0.54}_{0.39}$  & $2.13 \pm^{0.32}_{0.27}$ & $0.29\pm^{0.03}_{0.03}$ & $10.2\pm0.2$ \\
Metal Poor Disk & 255   & $13.53\pm^{0.32}_{0.41}$  & $1.15 \pm^{0.25}_{0.30}$ & $0.32\pm^{0.05}_{0.05}$ & $12.1\pm0.3$ \\
Thamnos         & 35    & $13.79\pm^{0.15}_{0.23}$  & $0.31 \pm^{0.24}_{0.15}$ & $0.17\pm^{0.10}_{0.07}$ & $13.4\pm0.3$ \\
Wukong/LMS-1    & 18    & $13.60\pm^{0.30}_{0.93}$  & $0.53 \pm^{0.98}_{0.31}$ & $0.50\pm^{0.32}_{0.23}$ & $12.9\pm1.3$ \\
Arjuna$^*$      & 17    & $11.63\pm^{0.90}_{0.74}$  & $0.52 \pm^{1.41}_{0.35}$ & $0.54\pm^{0.19}_{0.18}$ & $10.9\pm1.1$ \\
Sequoia         & 11    & $13.11\pm^{0.63}_{1.69}$  & $1.18 \pm^{1.79}_{0.77}$ & $0.48\pm^{0.32}_{0.28}$ & $11.6\pm2.4$ \\
Helmi Streams   & 8     & $12.08\pm^{1.31}_{1.87}$  & $2.17 \pm^{1.66}_{1.30}$ & $0.36\pm^{0.32}_{0.22}$ & $9.4\pm1.9$ \\
I'itoi$^*$     & 3     & $11.88\pm^{1.64}_{4.26}$  & $0.66 \pm^{1.84}_{0.48}$ & $0.54\pm^{0.32}_{0.36}$ & $10.4\pm3.2$
\enddata
\tablecomments{Summary of results of the star formation history fits, ordered by the number of MSTO members, $N$, that we use in the SFR fit.  The third and fourth column contains the maximum a posterior values for $\mu_\mathrm{SFH}$ and $\sigma_\mathrm{SFH}$ respectively.  The fifth column lists $f_\mathrm{out}$, the membership fraction assigned to the outlier SFH component.  The final column lists the lookback time at which each substructure had assembled 84\% of its mass.\label{table:sfr parameters}}
\tablenotetext{*}{For Arjuna and I'itoi we adopt a scale invariant prior  instead of the default uniform prior on $\sigma_\mathrm{SFH}$ (See Section \ref{subsec:sfh fitting} and Appendix \ref{app:sfh fitting}).}
\end{deluxetable*}


Figure \ref{fig:n age hists} shows each substructure's measured age distribution, with the counts normalized and corrected for the selection function.  The substructures are ordered by the number of the MSTOs we used to fit their SFH.  GSE is by far the largest component of the metal poor halo and by far the most well represented in our MSTO sample, so we will use it as a baseline to which we compare all other substructures.  The age distribution of GSE is plotted in the background of every other substructure's panel with an arbitrary normalization to leave the subplots readable.  

Figure \ref{fig:n amrs} shows each substructure in age---metallicity space, with a matching layout to Figure \ref{fig:n age hists}.  Again for comparison, the age---metallicity relation (AMR) of GSE is plotted in the background of each panel as a shaded region, with the dark center line indicating the median weighted age in each metallicity bin, and the lighter band indicating the 16th and 84th percentile ages.  The median, 16th and 84th percentile ages are used for the other substructures as well, plotted instead as individual points with error bars.  Metallicity bins that only contain a single star are plotted as open circles without error bars.

Table \ref{table:sfr parameters} summarizes our SFH fits for each substructure.  The second column lists the size of the MSTO sample used in the fit.  The third and fourth columns list the best fitting mean and standard deviation of the Gaussian SFH.  The fifth column is the membership fraction $f_\mathrm{out}$ assigned to the outlier component.  The associated uncertainties are the 16th and 84th percentile values of the marginalized parameter posteriors.  The last column lists $\tau_{84}$, defined such that integrating the SFH from 14 Gyr to the time $\tau_{84}$ contains 84\% of the total distribution.  In other words, $\tau_{84}$ is that time at which the SFH has assembled 84\% of its total mass.  We offer this quantity as an estimate of when a substructure's star formation begins to truncate, having already assembled the majority of its mass.  This quantity is only a heuristic, but we find it to be useful nonetheless because the small sample sizes limit us to simple Gaussian shapes for the SFHs, and this statistic gives some natural estimate for the `edge' of a continuous Gaussian distribution.

\begin{figure*}[t!]
\epsscale{0.8}
\plotone{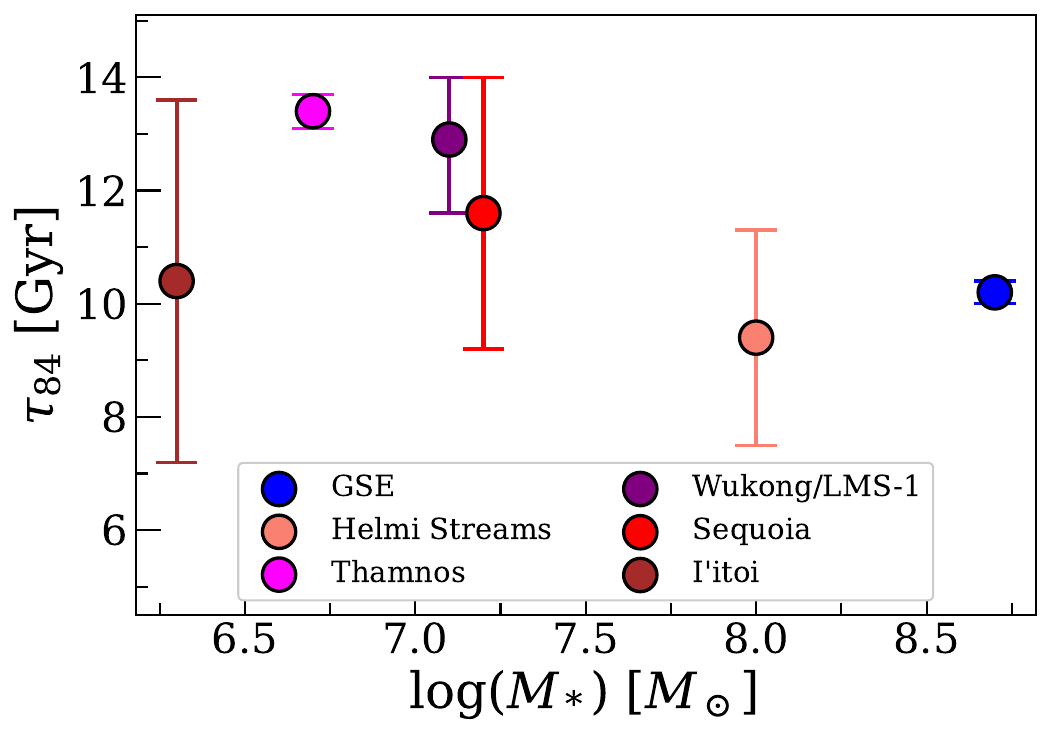}
\caption{The $\tau_{84}$ statistic from Table \ref{table:sfr parameters} compared to stellar masses estimates for each substructure.  The masses come from Table 1 in \citet{Naidu_2022b} (and references therein).  There is perhaps a slight trend, with the less massive substructures having truncated star formation at earlier times compared to the more massive systems like GSE and the Helmi Streams.  But we emphasize that the existence of any trend is very tentative and further work exploring this is needed.\label{fig:tau84 compare}}
\end{figure*}

Figure \ref{fig:tau84 compare} plots $\tau_{84}$ against the stellar masses of GSE, the Helmi Streams, Thamnos, Wukong/LMS-1, Sequoia, and I'itoi respectively.  The masses are those from \citet{Naidu_2022b}, who compiled dynamically estimated masses for GSE and the Helmi Streams \citep{Naidu_2021, Koppelman_2019b} and used relative star counts to estimate the masses of the other structures.  We tentatively observe a trend in this figure, with the lower mass systems tending to have shut off star formation at earlier times compared to the more massive systems.  However, further work is necessary to truly confirm the existence of this trend, and if so to then explain the physical mechanism causing the trend.  

Figure \ref{fig:1 panel SFH} displays the best fit star formation history for all eight substructures alongside each other, plotted with arbitrary normalizations and vertical offsets to leave the plot readable.  We plot with arbitrary normalization because we make no estimate on the absolute star formation rate, or equivalently, on the total mass of each system.  In this work we instead are only concerned with the \emph{relative} SFH within each substructure as a function of time, investigating how quickly and at what time each substructure assembled.  Arjuna and I'itoi's SFHs are plotted with hatches up and to the right to indicate that their SFH fits were less informative and required a stronger prior on $\sigma_\mathrm{SFH}$.  The MPD is plotted with hatches up and to the left because this is not a distinct dwarf galaxy or even necessarily a distinct subpopulation, but it is instead one slice of the larger star formation history of the Milky Way.

GSE is by far the largest component of the halo found within our sample so it is important to confirm that the other substructures, already distinguished in chemo-dynamical space, are also distinguishable from GSE in age space.  We test for contamination by adding the best fit SFH for GSE as an additional term into Equation \ref{eq:expanded sfh}, weighted by its own membership fraction $B^\prime$.  Relabeling the membership fraction of the primary structure as $A^\prime$, the resulting function fit is then $\psi(t|\theta) = A^\prime\psi_\mathrm{prim}(t|\theta) + B^\prime\psi_\mathrm{GSE}(t) + (1-A^\prime-B^\prime)\psi_\mathrm{out}(t)$.  If a given substructure's MSTO sample is heavily contaminated by GSE, then we can expect this fit to prefer large values of $B^\prime$ and small values of $A^\prime$ or prefer smearing out $\psi_\mathrm{prim}$ with large values for $\sigma_\mathrm{SFH}$, essentially only needing GSE's SFH to explain the observed age distribution.  By contrast if $A^\prime$ is large and the fit prefers a strong and distinct $\psi_\mathrm{prim}$, then we can be confident that we are indeed learning something about the SFH of that substructure and that it is a population distinct from GSE. 

When we perform this test the metal poor disk, Thamnos, and Wukong/LMS-1 each have marginal $B^\prime \lesssim 0.1$, $A^\prime > B^\prime$, and parameters $\mu_\mathrm{SFH}$ and $\sigma_\mathrm{SFH}$ fully consistent with the fiducial fits listed in Table \ref{table:sfr parameters}.  We can therefore be confident that these samples have very low contamination from GSE and that the age distributions and SFHs we measure are an accurate reflection of these populations.

The contamination tests for the Helmi Streams and Sequoia are more borderline, with marginal $A^\prime \approx B^\prime$.  We cannot know whether this is due to our already small samples having a nontrivial amount of contamination from GSE, or because the true SFH of these structures is close to that of GSE and they cannot be easily distinguished in age space alone.  In Sections \ref{subsubsec:HelmiStreams} and \ref{subsubsec:Sequoia} we discuss the implications of each of these two scenarios.  

The corner plot of Arjuna's contamination fit is shown in Figure \ref{fig:Arjuna_GSEcontamination} in Appendix \ref{app:sfh fitting}.  The result is bimodal in $\mu_\mathrm{SFH}$, with one mode having $A^\prime > B^\prime$ and corresponding to the same solution as was found in the fiducial fit.  But the second mode has $A^\prime\approx0.1$ and $B^\prime \approx 0.5$, indicating that GSE's SFH is sufficient to explain the majority of Arjuna's age distribution.  Section \ref{subsubsec:Arjuna} discusses the implications of this result to the origins of Arjuna.

Lastly, the fiducial fit for I'itoi is already barely informative due to the sample size of only 3 MSTOs, so it is not surprising that it fails the contamination test too.  The marginal posteriors of $A^\prime$ and $B^\prime$ are wholly unconstrained and so we cannot draw any conclusions about potential contamination from or connection to GSE.  Building a larger sample for I'itoi in the future is necessary to draw any further conclusions on the origin of this structure.

\section{Discussion}
\label{section:discussion}
\subsection{Caveats \& Limitations}
\label{subsec:caveats}
In this section we discuss limitations of these analyses and directions for future work.

Following \citet{Naidu_2020} we employed additional metallicity cuts to define our Thamnos, Wukong/LMS-1, Arjuna, Sequoia, and I'itoi MSTO samples.  These cuts serve to differentiate between substructures that have similar or overlapping kinematics.  However, if there exists a trend between age and metallicity in these substructures (as is the case for GSE, see Figure \ref{fig:n amrs}) then a metallicity cut will also impose a cut on the age distributions in Figure \ref{fig:n age hists}, biasing our results.  Using a purely kinematically definition for these substructures would avoid this problem, but it would also require a much more careful and complicated treatment of contamination e.g., a probabilistic membership model.  We do not attempt this here and instead opt for a more general overview of these substructures in age space.  We therefore cannot make any strong conclusions about the existence of an age---metallicity in any substructures, save for GSE and the metal poor disk which are selected without any additional cuts in abundance space, and where we have a sufficient number of stars over an extensive range of metallicities.  

The H3 survey's magnitude criterion limits MSTOs to be observed at distances between $\approx1.5-6$ kpc.  Our sample therefore misses several distant merger remnants known to be observable among the H3 giants, such as Sagittarius, Cetus, and Orphan/Chenab \citep{Naidu_2021}.  Without visible MSTOs in H3 we cannot offer any insight into the age structure of these progenitor galaxies.  Fortunately these three structures are still spatially coherent and so are more easily the subjects of targeted studies (e.g., \citealt{de_Boer_2015}, \citealt{Hansen_2018} for Sagittarius).

The magnitude criterion is also the primary reason for the small sample size of most substructures, with GSE and the MPD being the only exceptions.  The other substructures already have fewer stars compared to GSE owing to their lower masses, but the volume limitation when selecting for MSTOs further cuts down the number of usable stars.  So there is still plenty of room for a more focused investigation into the age structure of these lower mass systems.

H3's exclusion of fields in the galactic plane also causes us to miss any substructures that have more disk-like kinematics, or those that lie deeper in the center of the galaxy e.g., Kraken \citep{Kruijssen_2018}.  Therefore we do not claim to have a \emph{complete} census of the metal poor halo.  Instead we provide an in depth view into the age structure of GSE, and simply give an overview of the age structure and star formation histories of several other major components of the halo. 

When discussing each substructure individually we consider the scenario where the edge of the age distribution and SFH truncation time $\tau_{84}$ is synonymous with the time that the dwarf galaxy merged i.e., its star formation quenched because of the loss of gas during the merger.  This scenario is only an assumption; it is possible for star formation in dwarf galaxies to quench for other reasons e.g., stellar feedback, reionization, or environmental preprocessing \citep{Samuel_2022}.  But there is evidence from elemental abundances and merger simulations that this assumption is likely true for these specific substructures \citep{Naidu_2022b, Panithanpaisal_2021}.  

Also, we must note that there is some subtlety in what is meant by `merger time'.  The time between when a satellite galaxy initially accretes into the Galaxy's virial halo and when it is fully disrupted can vary depending on the Galaxy's mass at the time of accretion, the satellite's own mass, and the configuration of the satellite's orbit and infall.  For example, dynamical friction is less efficient on less massive satellites and therefore take longer to sink to the center of the host galaxy where they can experience stronger tidal forces.  So even when we are assuming that $\tau_{84}$ can be taken as a proxy for when a substructure merged, we are only assuming \emph{some} unspecified time between the initial accretion and final disruption \citep{Panithanpaisal_2021}.  

Lastly, we again emphasize that in this work we focus only on the metal poor components.  Most of the metal rich structures (e.g., the in situ halo/Splash, high$-\alpha$/thick disk, Aleph) have already been the subject of focused studies \citep{Bonaca_2017, Gallart_2019, Bonaca_2020, Belokurov_2020, Naidu_2020, Xiang_2022}, and though there remains conflicting conclusions between some of these studies, reconciling those conflicts is well beyond the scope of this work.

\subsection{Individual Substructures}
\label{subsec:comparison to literature}
In this section we discuss our results for each substructure individually, and place these results in the context of previous work timing the mergers of these various progenitors.  We compare to other reported stellar ages where available, but this work includes some of the first reported individual stellar ages for the Helmi Streams, Thamnos, Wukong/LMS-1, Arjuna, and I'itoi.  In these cases we must compare our results against other timing methods such as dynamical modelling or entire CMD fitting.

\begin{figure*}[t!]
\epsscale{1.0}
\plotone{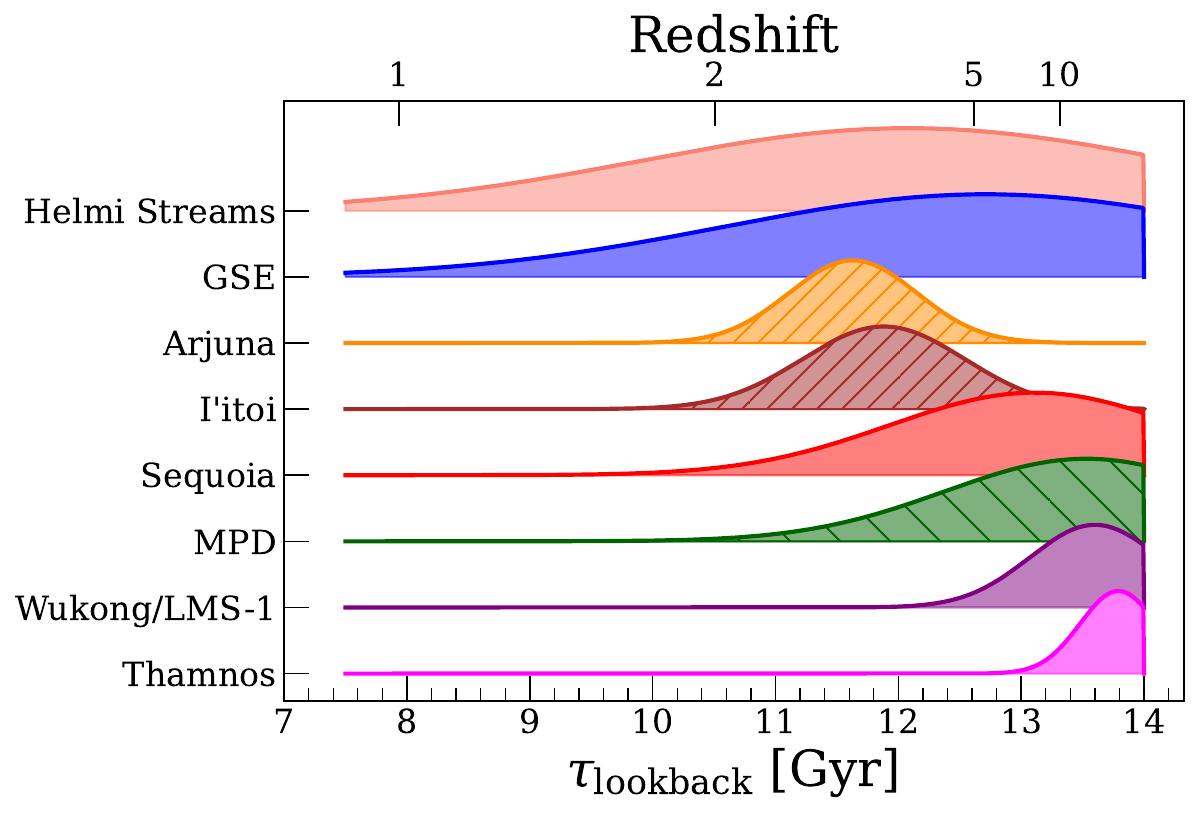}
\caption{Best fit star formation rates as a function of lookback time.  All SFHs are normalized and vertically offset to facilitate comparison, and ordered by the time of the distribution's 84th percentile ($\tau_{84}$, see Table \ref{table:sfr parameters}).  Three structures come with caveats and are hatched:  Arjuna and I'itoi required alternative priors in their fit (see Section \ref{subsec:sfh fitting}), and the metal poor disk (MPD) is only a slice of the Galaxy's larger in situ formation history, so these structures' SFHs cannot be interpreted so directly.  The Helmi Streams have an extended star formation history and assembled 84\% of their mass most recently.  On the other extreme, Thamnos formed very early and quickly, assembling all of its mass within only $\lesssim1$ Gyr.\label{fig:1 panel SFH}}
\end{figure*}

\subsubsection{GSE}
\label{subsubsec:GSE}

GSE is by far the most well studied of the eight structures we consider, with its merger time having been estimated across a variety of studies and methods.  These range from direct methods applied to GSE member stars or indirect methods that assume a causal connection between the merger and transformations in another populations.  Specific methods used include chemical evolution modelling \citep{Vincenzo_2019, Spitoni_2019, Spitoni_2020}, matching $N-$body or cosmological zoom$-$in simulations to observed dynamics and chemistry \citep{Belokurov_2018, Mackereth_2018, Naidu_2021}, dating the merger's chemical and kinematic imprint on in situ populations \citep{Di_Matteo_2019b, Ciuca_2023}, the qualitative placement or quantitative fitting of isochrones to a CMD \citep{Helmi_2018, Sahlholdt_2019, Gallart_2019}, globular cluster age--metallicity relations \citep{Myeong_2018b, Kruijssen_2018, Massari_2019, Forbes_2020}, or age dating individual stars via techniques such as chemical abundance mapping, isochrone fitting \citep{Das_2020, Bonaca_2020, Feuillet_2021, Xiang_2022}, or asteroseismology \citep{Chaplin_2020, Grunblatt_2021, Montalban_2021, Borre_2022}.

Estimates for the time of the merger range from $8-12$ Gyr ago, with a majority of the literature converging to a time of $9-11$ Gyr ago.  Indirect methods that use in situ populations often benefit from much larger sample sizes because in the solar neighborhood disk stars are much more numerous than halo stars, but they lack the ability to resolve any of the age structure within GSE itself.  Direct methods such as chemical evolution modelling and age dating individual stars therefore are the only ways to probe the evolution of accreted galaxies like GSE \emph{before} they merged.  There are several studies that have already done this, but there are still discrepancies between the age distributions they infer for GSE.  These differences primarily manifest across different age dating methods.  For example our distribution broadly matches other distributions measured through isochrone fitting \citep{Das_2020, Bonaca_2020, Feuillet_2021, Xiang_2022} with a range of ages from $\approx9-14$ Gyr and a mean/median around $\approx 12$ Gyr.  Studies calculating asteroseismic ages for GSE members have tended towards younger ages by $2-4$ Gyr, for example \citet{Grunblatt_2021}, \citet{Montalban_2021}, and \citet{Borre_2022} find the mean age of GSE stars to be $8\pm3$ Gyr, $9.7\pm0.6$ Gyr, and $9.5\pm1.3$ Gyr respectively.   While these slight discrepancies could be driven by systematics in seismic scaling relations \citep[see][]{Epstein_2014}, these asteroseismic studies are still severely limited in sample size ($N\lesssim10$), so this apparent tension between the seismic and isochrone ages could resolve itself in the future with missions like PLATO \citep{Rauer_2014} that promise much larger samples of metal poor stars.

Isochrone studies are not without their own issues and systematics \citep[Appendix A; but see also][]{Huber_2024, Nataf_2024} but there already exists sufficient astrometric and spectrophotometric data for them to achieve far larger samples.  Our own analysis calculates individual stellar ages for 1066 GSE MSTOs from which we measure the full age distribution of GSE directly.  The apparent drop in star counts at an age of $\approx9.5$ Gyr is in good agreement with the merger timing in the literature.  Alternatively we can look at our SFH fits of GSE:  the fiducial Gaussian shape assembles 84\% of its mass by a lookback time of $10.2\pm0.2$ Gyr.  When we instead adopt a uniform SFH distribution (see Appendix \ref{app:sfh fitting}) we find a similar statistic of $\tau_{84}=10.2$ Gyr with all star formation ending by an age of $t_1 = 9.4$ Gyr.  Again, this is fully consistent with the consensus time for GSE's merger.  

This 84\% assembly time matches the truncation time for GSE found in \citet{Bonaca_2020}, who also used H3 data but instead assumed a uniform distribution for the shape of GSE's star formation history.  Our GSE sample here is $\times2.5$ larger than that used by \citet{Bonaca_2020} and our abundance cuts allow the sample to extend to slightly higher metallicities (\fehi$\lesssim-0.4$) than they allowed (\fehi$\leq-0.6$).  The age---metallicity relation we measure in GSE then explains why we find GSE stars down to slightly younger ages than they do.  We can also note that the age---metallicity relation is broadly consistent with those inferred from GSE's globular clusters \citep{Massari_2019, Kruijssen_2018, Kruijssen_2020}, chemical evolution modelling \citep{Vincenzo_2019}, and other studies with individual age dating \citep{Xiang_2022}.

\subsubsection{Helmi Streams}
\label{subsubsec:HelmiStreams}
Despite our sample having only eight MSTOs belonging to the Helmi Streams, we can clearly see that it reaches a wider range of ages than most of the other accreted substructures (Figure \ref{fig:n age hists}).  The age range of our Helmi Streams sample reaches as young as $\approx9$ Gyr, similar to that of GSE, and our SFH fit measures 84\% of the mass assembled $\tau_{84}=9.4\pm1.4$ Gyr ago.  This suggests that the Helmi Streams had a more extended star formation history than the majority of the other disrupted dwarfs we consider, broadly consistent with a fairly massive and evolved progenitor expected by the Streams' dynamics and abundances \citep{Koppelman_2019b, Naidu_2020, Horta_2022}.  Again we are assuming that the SFH truncation time is synonymous with the progenitor's merger time.  The specific merger time we infer loosely agrees with the old end of the range of 5---8 Gyr ago estimated dynamically by \citet{Koppelman_2019b}.  So despite having only a handful of stellar ages our results are generally consistent with other estimates of the age structure \citep{Ruiz_lara_2022} and merger, with the Helmi Streams' progenitor likely merging at a time comparable to or slightly after GSE.

\subsubsection{Thamnos}
Following \citet{Naidu_2020} we have selected a total of 35 MSTOs belonging to Thamnos and virtually all of them are older than 12 Gyr, with a very prominent overabundance of stars with ages $\approx13$ Gyr (Figure \ref{fig:n age hists}).  Only a single star satisfying the Thamnos selection is younger than 12 Gyr, with an age of $\approx9.5$ Gyr, so we find that this star is could be an outlier from another population.  The SFH we fit is incredibly compact in time, with the distribution's 84th percentile at an age of $\tau_{84}=13.4\pm0.3$ Gyr.  We again have to note that our absolute age scale is imperfect so we cannot safely convert such old ages or early times into corresponding redshift values.  Nevertheless, Thamnos is the oldest of all of the substructures we consider here.  Combined with the fact that the substructure is very deep in the Galaxy's gravitational potential and its overall very low metallicity \citep{Koppelman_2019a}, Thamnos is likely one of the earliest known contributions to the Milky Way's stellar halo.  This age distribution marks some of the first individual ages reported for any members of Thamnos, and the measured age distribution and SFH broadly agrees with that found through entire CMD fitting \citep{Dodd_2024}.

\subsubsection{Wukong/LMS-1}
We find 18 MSTOs as likely members of Wukong/LMS-1, and similar to Thamnos they are almost all very old with ages $>12$ Gyr.  Unlike our Thamnos sample however, the majority of our Wukong/LMS-1 MSTOs are inferred to have unphysically old ages, with a notable peak around 16 Gyr in Figure \ref{fig:n age hists}.  It is unclear what is causing this, especially considering that Wukong/LMS-1 and Thamnos have similar abundances but the ages of our Thamnos stars are much better behaved.  Keeping this caveat in mind, our ages and fitted star formation history suggest a lower limit of $\approx 12-13$ Gyr on the accretion time for Wukong/LMS-1.  

\citet{Malhan_2021} ran dynamical models of the Wukong/LMS-1 merger and found a most probable accretion time of $>8$ Gyr ago, a lower limit but still a time in slight tension with our estimate.  They go on to argue that the unusually low eccentricity and high velocity dispersion for stars accreted from a dwarf galaxy suggest that the progenitor was relatively massive and accreted at early times, with it undergoing significant orbit circularization before it was fully disrupted.  An early and drawn out accretion could explain the discrepancy between the dynamical and age based accretion times, but a discrepancy of up to 4 Gyrs could be too large and indicate an inconsistency between the two methods.  Some additional and independent age information could help to reinforce our finding. 

To that end, we can consider a number of globular clusters that have been chemo-dynamically associated with Wukong/LMS-1 by a variety of previous sources.  There is ample evidence for the association of the clusters NGC 5024 and NGC 5053 \citep{Yuan_2020b, Naidu_2020, Bonaca_2021, Malhan_2021, Malhan_2022a}, and there is tentative evidence for NGC 5272 \citep{Bonaca_2021, Malhan_2022a}, ESO 280-SC06 \citep{Naidu_2020}, and Palomar 5 \citep{Malhan_2021, Malhan_2022a}.  NGC 5024 and NGC 5053 have age estimates ranging from $\approx12.25-13.5$ Gyr \citep{Marin_Franch_2009, Dotter_2010, Forbes_2010, Vandenberg_2013}, consistent with our age distribution for Wukong/LMS-1.  NGC 5272 is slightly younger but still consistent, with age estimates ranging from $11.5-12.5$ Gyr \citep{Marin_Franch_2009, Dotter_2010, Forbes_2010, Vandenberg_2013}.  Palomar 5 on the other hand has age estimates ranging from $10-12$ Gyr and a mean metallicity of [Fe/H]$\approx-1.3$ \citep{Forbes_2010, Dotter_2011}, which is more metal rich than the abundance cut we use to select Wukong/LMS-1 members (\fehi$<-1.45$).  

If Palomar 5 genuinely originates from Wukong/LMS-1 then it would suggest that this metallicity cut is in fact too strict, and that we are using a biased sample to study this substructure.  It follows that our age distribution could also be biased to only include older and more metal poor members, potentially explaining the discrepancy between the accretion times inferred from dynamics and isochrone ages.  However, the streams Indus, Jhelum, Phoenix, Ravi, Sv{\"o}l, C-19, and Sylgr have each been associated to Wukong/LMS-1 by one or multiple studies, and all have metallicities in the range [Fe/H]$=-1.8$ to $-3.4$, well within our metallicity cut \citep[][and references therein]{Bonaca_2021, Malhan_2021, Malhan_2022a, Limberg_2024}.  That leaves Palomar 5 the sole outlier in terms of metallicity and so further work to confirm or reject its association with Wukong/LMS-1 could resolve whether our metallicity cut is too strict and biasing our age distribution.

\subsubsection{Arjuna}
\label{subsubsec:Arjuna}
Arjuna was first discovered by \citet{Naidu_2020} as an overabundance of halo stars with [Fe/H]$\approx-1.2$ on high energy retrograde orbits.  Using $N-$body simulations of the GSE merger \citet{Naidu_2021} later argued that Arjuna did not reflect a distinct galaxy merger but was in fact just debris from the outskirts of GSE that was stripped early on, before the progenitor's initially retrograde orbit had been significantly radialized.  Recent analysis of detailed elemental abundances also supports a common origin for Arjuna and GSE stars \citep{Horta_2022}.  

We have 17 MSTOs in our Arjuna sample and their age distribution reaches ages as young as 10 Gyr, several Gyr younger than most of the other substructures e.g., Thamnos, Wukong/LMS-1, and Sequoia (Figure \ref{fig:n age hists}).  Though it is possible that the progenitor of Arjuna simply accreted after these other systems and was therefore able to keep forming stars for longer, our SFH fits actually prefer the scenario where GSE is the progenitor of Arjuna.  As described in Section \ref{subsec:sfh fitting}, we tested how a substructure's SFH fit changes when we include GSE's SFH as an additional component.  When we run this test for Arjuna, it fails (see Appendix \ref{app:sfh fitting} and Figure \ref{fig:Arjuna_GSEcontamination}).  In other words, Arjuna's age distribution can be better described by using only GSE's SFH, rather than invoking a second distinct SFH.  So in addition to its kinematics and detailed abundances, our analysis proves that Arjuna is also indistinguishable from GSE in age space.  

We tentatively conclude that the GSE progenitor lacked a spatial age gradient on the basis of the similar age distributions of our Arjuna and GSE samples.  While this conclusion is limited by the small size of our Arjuna sample, it seems to be consistent with the findings of \citet{Naidu_2021} that GSE lacked a significant spatial metallicity gradient, despite the existence of an overall age---metallicity gradient.

\subsubsection{Sequoia}
\label{subsubsec:Sequoia}
Since its discovery \citep{Myeong_2019, Matsuno_2019} there has been a variety of speculation about possible origins for the Sequoia system.  It has been suggested that Sequoia is the high energy retrograde debris of GSE \citep{Koppelman_2019a, Koppelman_2020}, Sequoia reflects a distinct progenitor but one that was a satellite of GSE \citep{Myeong_2019, Matsuno_2019}, Sequoia and the lower energy retrograde structure Thamnos share a progenitor \citep[see discussion in][]{Monty_2020}, or whether it is simply its own distinct galaxy remnant that doesn't bear any association to other observed substructures \citep{Monty_2020, Aguado_2021, Garcia_Bethencourt_2023}.  The nature of Sequoia has remained ambiguous because of the difficulty inherent in selecting a pure and unbiased sample in chemo-dynamical space when Sequoia has significant overlap with several other structures e.g., Arjuna, I'itoi, Thamnos, and GSE \citep[see][]{Feuillet_2021, Horta_2022}.  Adding the additional information of stellar ages has the potential to clear up some of this ambiguity in the future, but at present we are limited by the size of our sample and so we will not make many strong conclusions.

All but one of the 11 Sequoia members have ages $>12$ Gyr, suggesting that Sequoia likely had a star formation history that was more similar to Thamnos or Wukong/LMS-1 rather than to GSE.  When the additional GSE component is included in the simple star formation history fit for Sequoia (see Section \ref{subsec:sfh fitting}) we find that Sequoia remains as its own term distinct from GSE, though we note that the fits for Sequoia are very uncertain with or without a GSE term due to using so few stars.  We therefore find it unlikely that Sequoia is debris from the same progenitor as GSE, and Arjuna is instead the more likely candidate to be the debris stripped from the outskirts of GSE onto retrograde orbits.  

Though the chemo-dynamics of Sequoia have been studied extensively, there are only a few studies that have inferred ages that we can compare against our own results.  \citet{Alencastro_Puls_2021} identify one or two metal poor stars with asteroseismic ages of $\approx10$ Gyr and $\approx14$ Gyr respectively that could potentially be members of Sequoia.  Another study by \citet{Feuillet_2021} used \emph{Gaia} DR2 RVS stars informed by APOGEE DR16 abundances to select both Sequoia and GSE populations, though their Sequoia population extends to much lower energies and may be mixed with the population we separate as Thamnos.  Fitting isochrones to the \emph{Gaia} DR2 photometry while conditioning on APOGEE abundances they found a similarly old age distribution for Sequoia that had a prominent peak of ages around $12-14$ Gyr, very consistent with our own inferred age distribution.  \citet{Dodd_2024} used full CMD fitting to measure Sequoia's SFH and found similar results.  Taking all of these results together, Sequoia appears to truncated a couple Gyr before GSE sample, reinforcing the scenario that Sequoia and GSE do not come from a common progenitor.

\subsubsection{I'itoi}
Out of all of the structures we consider in this work, I'itoi is by far the least well studied.  Originally discovered by \citet{Naidu_2020} as the most metal poor of the three peaks in the MDF of the high energy retrograde halo, this remnant reflects what is likely the least massive and least evolved progenitor in our entire sample \citep{Naidu_2022b}.  Owing to its subsequently low star count only three MSTOs are assigned to I'itoi, and our SFH fit essentially only reflects the mean of the three ages we infer.  Even if we assume all three of these stars are genuine I'itoi members then the most we can conclude is that I'itoi is generally old, which is to be expected from its very primitive abundances \citep{Horta_2022}.

\subsubsection{Metal Poor Disk}
\label{subsubsec:MPD}
The sample we label as the metal poor disk (MPD) does not necessarily represent debris from any merged galaxy but is instead part of the metal poor tail of the Milky Way's disk \citep[also known as the metal weak thick disk;][]{Carollo_2019}.  If the disk extends smoothly down to low metallicities then our MPD sample is simply the portion of the disk that happens to satisfy the abundance cuts adopted in Section \ref{subsec:substructure selection}, instead of being a meaningfully distinct subpopulation present within the disk.  In this case the young edge of the MPD SFH is a result of this abundance cut applied to the disk's age---metallicity gradient.  So instead of focusing on the apparent drop in star formation rate at younger ages, we emphasize the existence of a metal poor tail to the Milky Way's disk that extends to very old ages.  

It is clear from Figures \ref{fig:n age hists} \& \ref{fig:n amrs} that at a fixed metallicity the disk is older than GSE, consistent with expectations that the more massive Milky Way would enrich more quickly than GSE and reach a given metallicity at an earlier epoch.  These stars remain at ages $\gtrsim12$ Gyr up until a metallicity of \fehi$=-1.0$.  Here we cannot properly compare the number or total mass of these stars against their more metal rich counterpart the high$-\alpha$ disk, but it is clear that there exists a substantial ancient and metal poor tail to the Milky Way's stellar disk.

The footprint and selection of H3 limits our MPD MSTO sample to only contain stars with $R_{apo}\gtrsim 6$ kpc, and so it is not clear from our sample alone how this ancient metal poor disky population ties into the more centrally concentrated, metal poor spheroid that is the remnant of the protogalactic Milky Way (alternatively dubbed Aurora by \citealt{Belokurov_2022a} and the Poor Old Heart by \citealp{Rix_2022}).  \citet{Conroy_2022} demonstrates that the high$-\alpha$ disk does extend smoothly back into this spheroidal component, so it is possible that our particular MPD population is just the low eccentricity tail of the protogalactic spheroid.  Alternatively, the MPD could reflect only the very beginning of the proper, rotationally supported disk.  In either case, these nearby ancient stars offer a window into the chemical evolution of the Milky Way at its earliest epochs.

\section{Conclusions}
\label{section:conclusion}
Using data from the H3 survey we selected main sequence turn off and subgiant stars that are members of metal poor halo substructures as defined by \citet{Naidu_2020}. Using the Bayesian stellar isochrone fitting code \texttt{MINESweeper} we refit ages for this sample of $\approx2000$ total metal poor stars.  We used the resulting age distributions to fit simple star formation histories and infer the evolution and merger times for the progenitors of GSE, the Helmi Streams, Thamnos, Wukong/LMS-1, Sequoia, Arjuna, and I'itoi.  Additionally, we consider the remaining MSTOs with prograde disk-like orbits as a group that is part of the larger in situ disk population.  The age distributions, age---metallicity distributions, and best fit simple star formation histories can be found in Figures \ref{fig:n age hists}, \ref{fig:n amrs}, and \ref{fig:1 panel SFH} respectively.

\begin{enumerate}
\item We find that GSE has a relatively flat distribution of ages that extends from the time of the earliest stars $\sim14$ Gyr ago to 9.4 Gyr ago.  We interpret the distinct lack of younger stars as indication that GSE merged with the Milky Way 9.4 Gyr ago, in agreement with the consensus within the literature that the merger occurred $9-11$ Gyr ago.  We also resolve a clear age---metallicity relation within GSE, 
one that is flat at an age of 14 Gyr until a metallicity of \fehi$\approx-1.7$, after which metallicity increases linearly in time until its accretion and tidal disruption.

\item Despite a small sample size of eight MSTOs, the Helmi Streams clearly display a wide range of ages and therefore an extended star formation history, consistent with the picture that the progenitor of the Helmi Streams was a massive satellite galaxy \citep{Koppelman_2019b}.  We estimate a merger time of $9.4\pm1.9$ Gyr ago, broadly in agreement with other timing estimates that range from $5-11$ Gyr ago \citep{Koppelman_2019b, Kruijssen_2020}.  Though our uncertainties on this time are large, we find that the Helmi Streams merged at a time similar to or after GSE merged.

\item Virtually all of the 35 MSTOs we classify as members of Thamnos are very old with ages $>12$ Gyr.  These ages, their orbits deep in the Galactic potential, and the very low metallicity of Thamnos make it one of the most ancient contributions to our Galaxy's merger history.

\item We find that Wukong/LMS-1 and Sequoia are both very old as well, with the vast majority of their 18 \& 11 MSTOs respectively also being older than $12$ Gyr.  Interpreting the edge of their star formation histories as the time of their merging with the Milky Way, they both accreted as similar epoch $11-13$ Gyr ago, likely after Thamnos but before GSE.  We note that there is slight tension between our isochrone inferred accretion time and the dynamically inferred accretion time for Wukong/LMS-1 \citep{Malhan_2021}.  For Sequoia, though our SFH fits have large uncertainties we nevertheless find it to have a distinct SFH from GSE, meaning this substructure was not part of GSE and is instead the debris of a separate galaxy merger.  Due to the difference in their implied accretion time, we also find it unlikely that the Sequoia galaxy was itself a satellite of GSE, though our sample is too small to definitively rule out this possibility.

\item We find 17 MSTOs in Arjuna with an age distribution that appears to extend to slightly younger ages than Thamnos, Wukong/LMS-1, and Sequoia.  We find that Arjuna does not have a distinct SFH from GSE, supporting the conclusion of \citep{Naidu_2021} that Arjuna represents the most retrograde debris that was stripped from GSE at the earliest times during its merger.

\item We only find three MSTOs in I'itoi, and so the most we can surmise about this substructure is that it appears to be generally old with an average age of $\approx 12$ Gyr.

\item Lastly we find that the MPD is very ancient; its age distribution peaks at $\approx13$ Gyr and it exhibits an age---metallicity relation where it is older than GSE when compared at any fixed metallicity.  Though it is clearly some remnant of the ancient protogalaxy, it is not clear whether the MPD as we define it is the low eccentricity tail of the central protogalactic spheroid \citep{Belokurov_2022a, Rix_2022}, the oldest part of the Galactic disk \citep{Conroy_2022}, or more likely a combination of both.
\end{enumerate}

\newpage
\begin{acknowledgments}

CC and PC acknowledge support from NSF grant NSF AST-2107253.  We thank the Hectochelle operators and the CfA and U. Arizona TACs for their continued support of the H3 Survey. Observations reported here were obtained at the MMT Observatory, a joint facility of the Smithsonian Institution and the University of Arizona.  This paper uses data products produced by the OIR Telescope Data Center, supported by the Smithsonian Astrophysical Observatory. The computations in this paper were run on the FASRC Cannon cluster supported by the FAS Division of Science Research Computing Group at Harvard University.  

Support for this work was provided by NASA through the NASA Hubble Fellowship grant HST-HF2-51515.001-A awarded by the Space Telescope Science Institute, which is operated by the Association of Universities for Research in Astronomy, Incorporated, under NASA contract NAS5-26555. 

This work has made use of data from the European Space Agency (ESA) mission \emph{Gaia} (\url{https://www.cosmos.esa.int/gaia}), processed by the \emph{Gaia} Data Processing and Analysis Consortium (DPAC, \url{https://www.cosmos.esa.int/web/gaia/dpac/consortium}) \citep{Gaia_2021}. Funding for the DPAC has been provided by national institutions, in particular the institutions participating in the \emph{Gaia} Multilateral Agreement.  Funding for the Sloan Digital Sky Survey IV has been provided by the Alfred P. Sloan Foundation, the U.S.  Department of Energy Office of Science, and the Participating Institutions. SDSS-IV acknowledges support and resources from the Center for High Performance Computing at the University of Utah. The SDSS website is \url{www.sdss.org}.

\facility{
MMT (Hectochelle), \emph{Gaia}
}
\software{
\texttt{Astropy} \citep{astropy, astropy_2022}, \texttt{emcee} \citep{emcee}, \texttt{matplotlib} \citep{matplotlib}, \texttt{numpy} \citep{numpy}, \texttt{scipy} \citep{scipy}
}

\end{acknowledgments}

\appendix
\section{Synthetic Age Recovery}
\label{app:isochrone mocks}
Isochrone ages are most precise around the main sequence turn off and subgiant branch where the isochrones are most widely separated on a CMD or Kiel diagram.  However, these diagrams are just projections of the entire high dimensional space of stellar isochrone fitting.  As mentioned in Section \ref{subsec:isochrone ages}, priors can be placed on any of the 14 primary parameters sampled by \texttt{MINESweeper}, as well as any further parameters derived from them.  The physics of stellar evolution and atmospheres is complicated and highly degenerate, so seemingly simple priors on one parameter can propagate into unexpected and non-trivial priors on other parameters.  This means that the inference of particularly sensitive parameters, namely age, can easily be biased during the inference.  

We test age recovery in our own fitting procedure by creating synthetic mock observations, fitting them under the same system of priors used to fit the data, and checking how successfully the input age is recovered.  We test how age recovery varies with log(g) (a proxy for evolutionary phase), S/N ratios, and intrinsic properties age and \fehi.  We also explore some alternative priors and assumptions to see their effect on the age inference e.g., neglecting atomic diffusion or fitting an unresolved binary with a single star solution.

\begin{figure*}
\epsscale{0.8}
\plotone{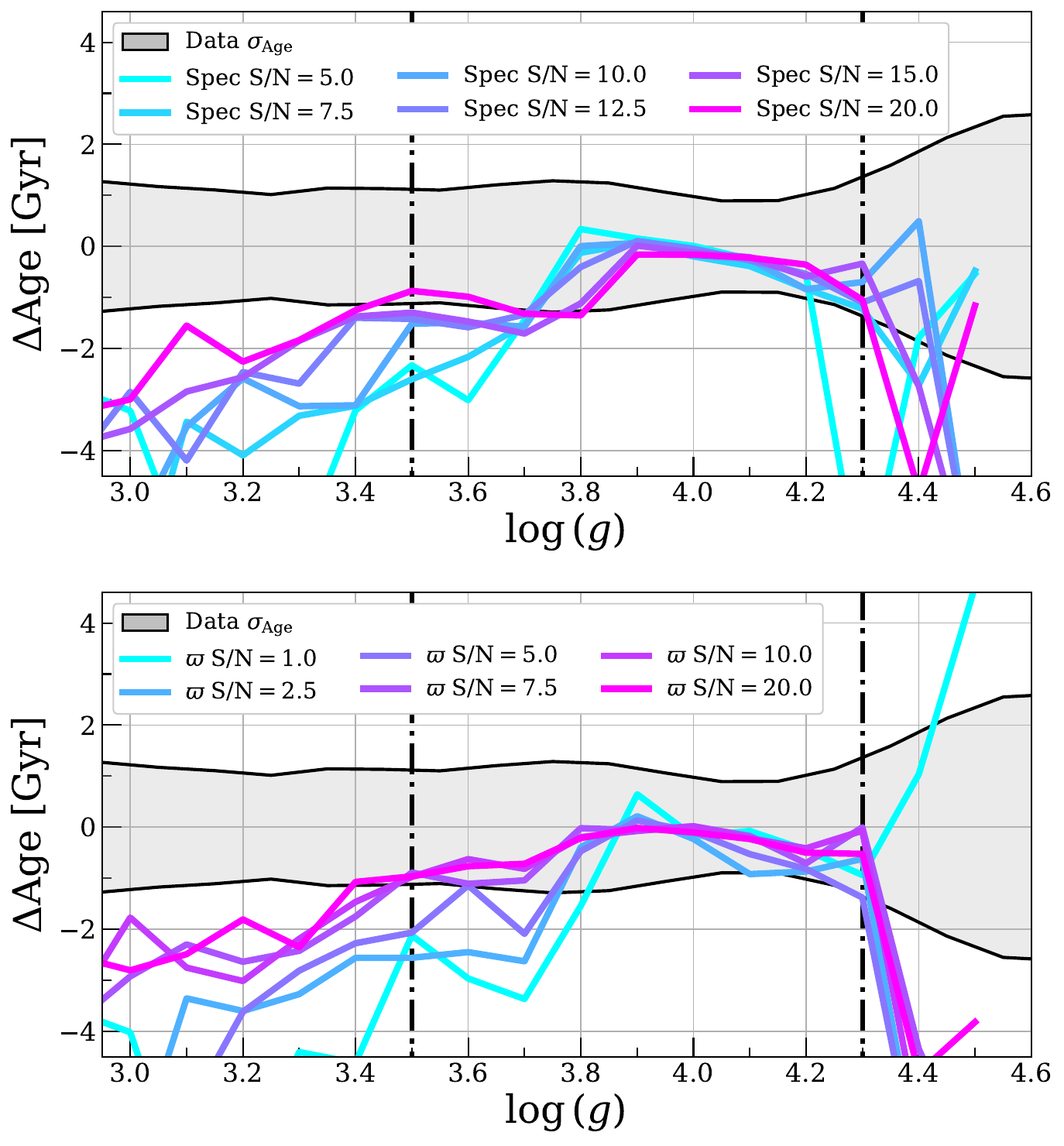}
\caption{Age recovery as a function of \logg\ at different spectral and parallax S/N ratios.  The data's median age uncertainty as a function of \logg\ is plotted as a mirrored black line and shaded region.  Mock observations were generated with an age of 10 Gyr, \fehi$=-1.0$, \afei$=0.2$, Distance$=3$ kpc, and \Av$=0.1$ in all cases.  A fixed parallax S/N $=5$ was used in the top panel and a fixed spectral S/N $=10$ was used in the bottom panel.  Ages can only be reliably recovered for \logg\ values around the main sequence turn off and subgiant branch, so we only select stars with $3.5<$ \logg\ $<4.3$, indicated by the two vertical dotted lines.\label{fig:snr mocks}}
\end{figure*}

\begin{figure*}
\epsscale{0.8}
\plotone{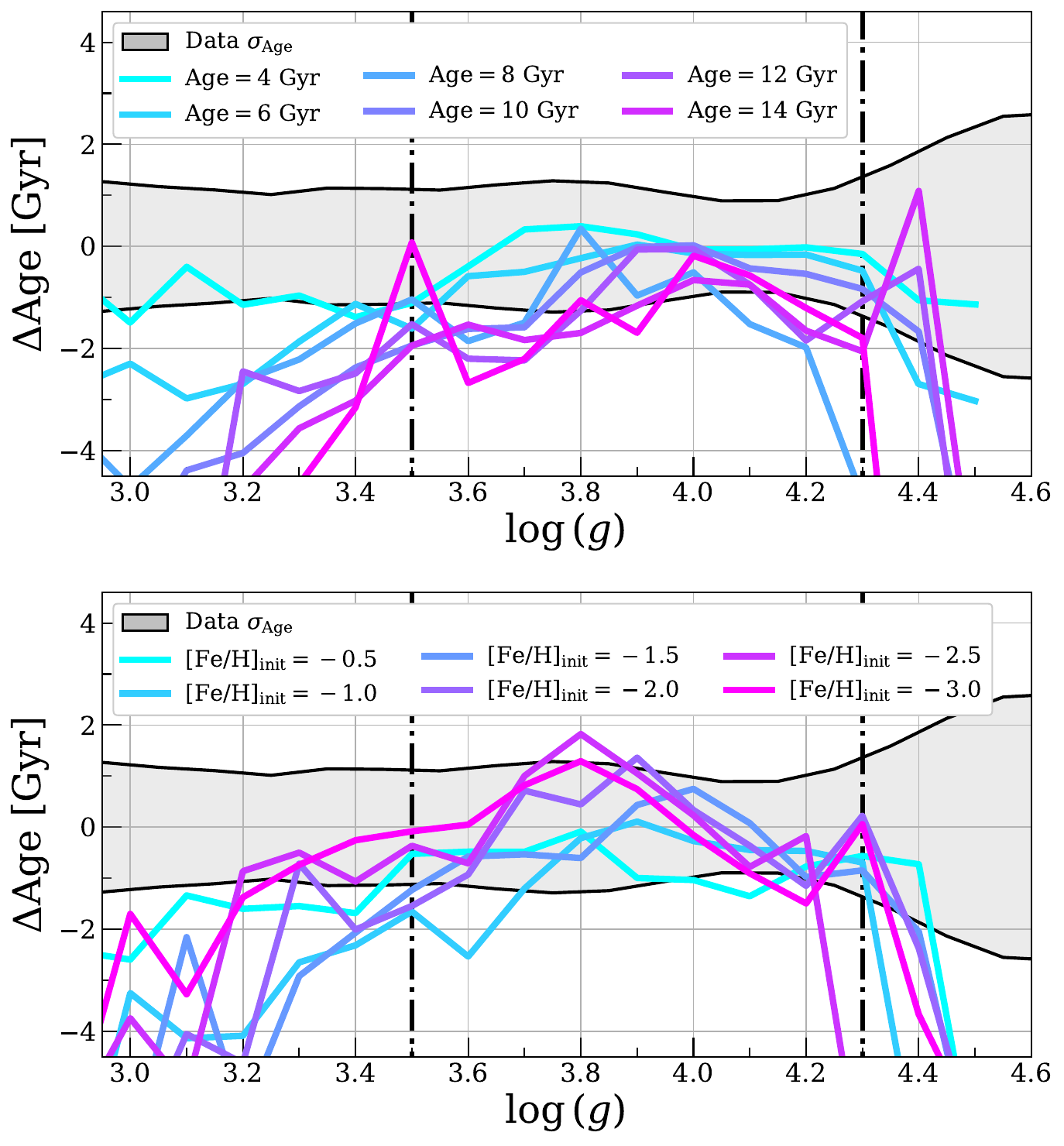}
\caption{Age recovery as a function of \logg\ at different ages and initial metallicities, plotted in the same format as Figure \ref{fig:snr mocks}.  These mocks observations were generated with the same parameters as those in Figure \ref{fig:snr mocks} unless otherwise indicated by the legend.  In each case we set spectral S/N$=10$ and parallax S/N$=5$.  The age recovery is only marginally dependent on metallicity or age, therefore it is safe to use a single \logg\ cut to define our MSTO sample. \label{fig:age mocks}}
\end{figure*}

\begin{figure*}
\epsscale{0.8}
\plotone{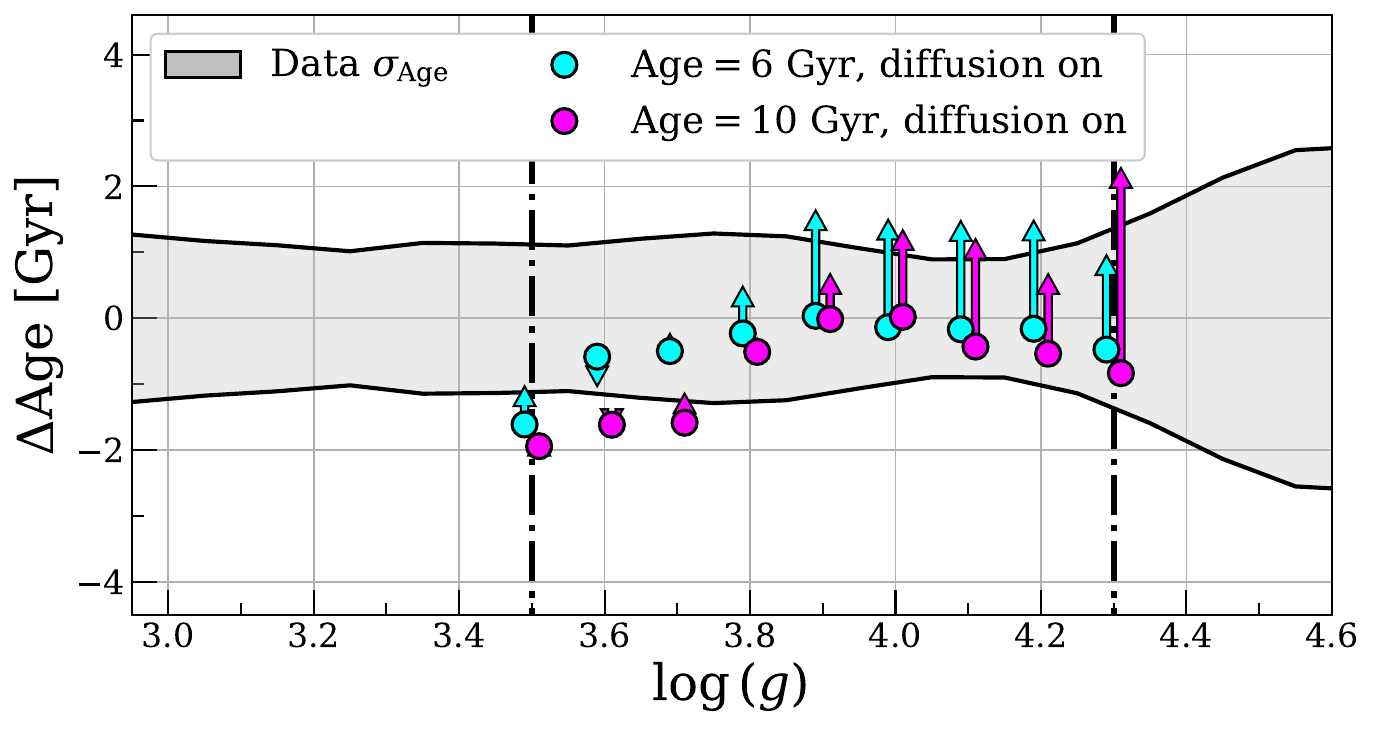}
\caption{Age recovery as a function of \logg\ plotted in the same format as Figure \ref{fig:snr mocks}, with the arrows showing what happens to the age recovery when the effects of atomic diffusion are ignored when fitting.  These mocks observations were generated with the same parameters as those in Figure \ref{fig:snr mocks} unless otherwise indicated by the legend.  In each case we set spectral S/N$=10$ and parallax S/N$=5$.  Ignoring atomic diffusion tends to bias ages to be older than they truly are for stars with higher \logg.  Stars at higher \logg\ are also more numerous, so ignoring diffusion will introduce an age bias to the majority of stars in an MSTO selected sample.\label{fig:diffusion mocks}}
\end{figure*}

\begin{figure*}
\epsscale{0.8}
\plotone{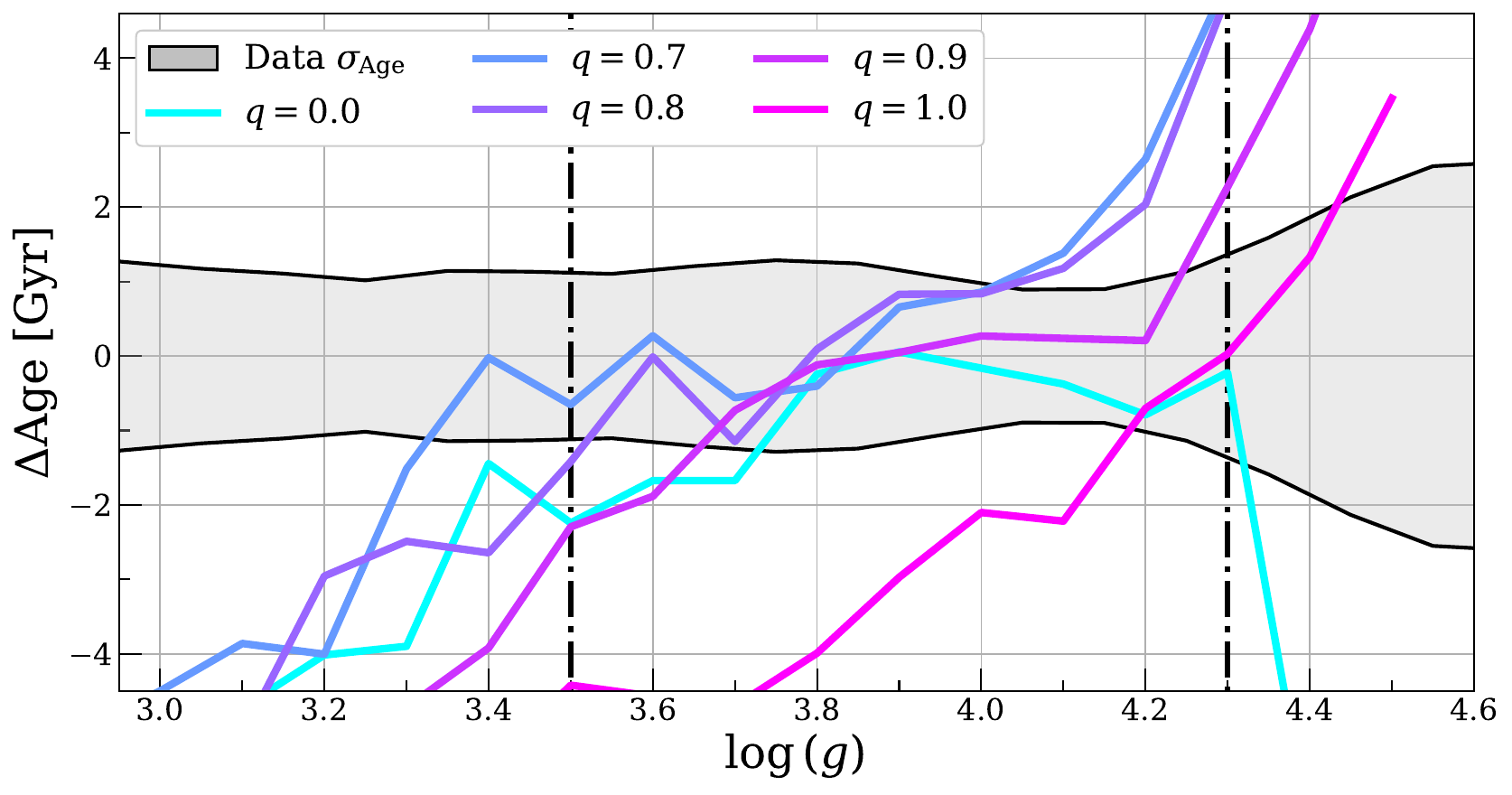}
\caption{Age recovery as a function of \logg\ for a set of unresolved binaries with varying secondary mass fraction $q$, plotted in the same format as Figure \ref{fig:snr mocks}.  These mocks observations were generated with the same parameters as those in Figure \ref{fig:snr mocks} unless otherwise indicated by the legend, and in each case we set spectral S/N$=10$ and parallax S/N$=5$.  The presence of unresolved binaries affects age recovery in a non-trivial way, with unequal mass binaries generally being biased to older ages, and equal mass binaries being significantly biased to younger ages.  \label{fig:binary mocks}}
\end{figure*}

Figure \ref{fig:snr mocks} shows the age recovery as a function of \logg\ at different spectral and parallax S/N.  In most cases the age recovery is worse than the typical statistical age uncertainty at the same \logg, meaning that age inferences can actually be dominated by systematic biases.  Only around the main sequence turn off and subgiant branch is the age recovery good enough for inferred ages to be reliable given the spectral and parallax S/N typical of H3 stars.  The age recovery appears to be only slightly sensitive to spectral S/N, but it is noticeably sensitive to parallax S/N.  We therefore adopt requirements for both spectral and parallax S/N to ensure that the age recovery is comparable to the age's statistical uncertainty.

Figure \ref{fig:age mocks} shows the results when we keep the S/N ratios fixed and instead vary the age and initial metallicity of the mocks.  Again, only stars around the turn off and subgiant branch have their ages reliably recovered.  We can additionally see that during these phases there is no substantial difference in age recovery across different ages and metallicities.  We can therefore apply the same \logg\ cut to all of our stars independent of their physical properties.  Together these recovery experiments inform the selection windows of $3.5<$ \logg\ $<4.3$, spectral S/N $>7$, and parallax S/N $>5$ that we adopt in Section \ref{section:data}, ensuring that our ages and SFH results are not being biased.  

Figure \ref{fig:diffusion mocks} shows the results when we turn off atomic diffusion during the fitting procedure i.e., inputting \feh\ instead of \fehi\ to the isochrone models.  Surface abundances become depleted around the main sequence turn off and are gradually restored to initial levels as a star moves toward the giant branch and its convective envelope deepens.  Neglecting diffusion and assuming that the initial abundance \fehi\ is equal to the depleted surface abundance \feh\ means using isochrones that are too metal poor and bluer than the star truly is.  So the isochrone fitter must then use older, redder isochrones to better match the observed SED and surface temperature.  This explains why our mocks fitted while neglecting the effects of diffusion are biased to ages $1-2$ Gyr older around the MSTO.  This systematic is compounded in an actual data set because the number density of stars increases with increasing \logg, making the majority of stars in an MSTO selected sample sensitive to this effect.

Figure \ref{fig:binary mocks} shows the results when we include the light from an unresolved binary companion to the spectrum and photometry of the mock observation.  The effect of binaries is complicated.  Equal mass binaries have identical spectra and colors and are simply twice as bright as their geometric parallax would suggest, and so they are fit to be more intrinsically luminous and therefore younger.  The effect of unequal mass ratios is most prominent at higher \logg\ where the light ratio between the primary and secondary is more balanced.  These binaries will then generally appear redder and slightly brighter, which translates into appearing older.  This effect gets dramatically worse at higher \logg\  because on the main sequence small changes in color imply several Gyr of evolution.  Our tests here are admittedly simple, and there is room for further work to model a realistic binary to see how the inferred age distribution and star formation history will be affected.

Our adopted \logg\ selection is an effort to balance this binary bias, the general age recovery bias demonstrated in Figures \ref{fig:snr mocks} \& \ref{fig:age mocks}, and the statistical uncertainty of inferred ages (see Figure \ref{fig:kielposteriors}) with the decreasing number density of stars at later evolutionary phases.

\section{Alternative SFH Fits}
\label{app:sfh fitting}
\begin{figure*}
\epsscale{1.0}
\plotone{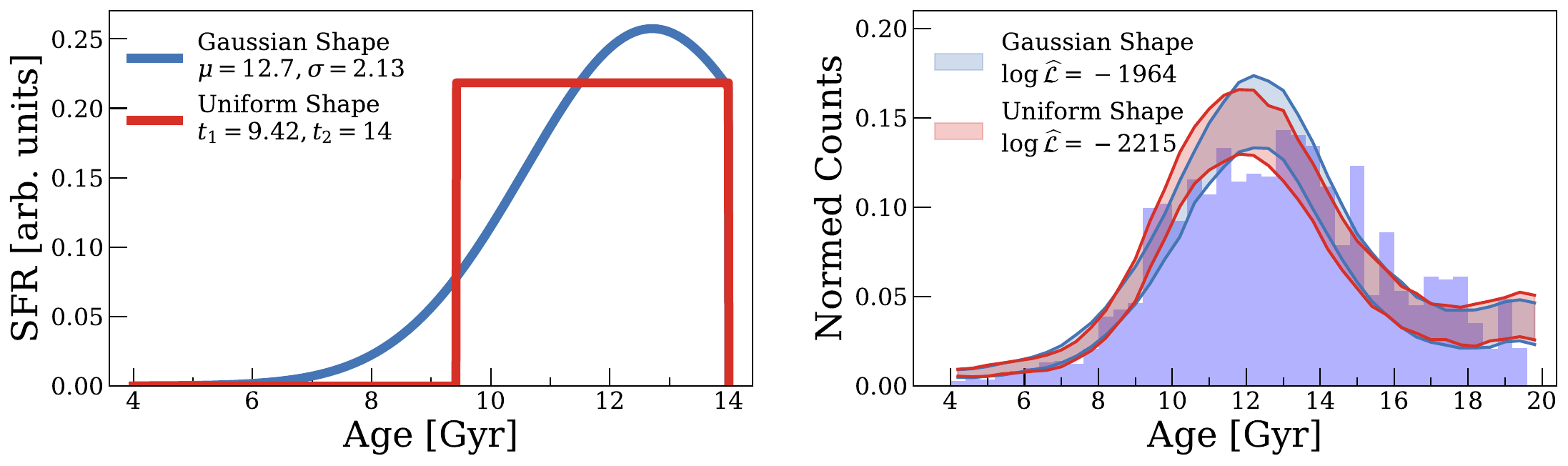}
\caption{\emph{Left}: The best fit analytic model for GSE's SFH assuming a Gaussian or uniform shape.  \emph{Right}: Error convolved and posterior sampled SFH fits for the Gaussian and uniform shapes.  The differences between the two shapes is slight, with their max likelihoods indicating that the Gaussian shape is slightly preferred.  Both the uniform and Gaussian shapes have $\tau_{84}=10.1-10.2$ Gyr.\label{fig:GSE_2shapes}}
\end{figure*}

GSE is by far the most well represented substructure in our MSTO sample.  With its $>1000$ stars we can meaningfully differentiate between different shapes for its star formation history.  Figure \ref{fig:GSE_2shapes} compares the results of our fits using fiducial Gaussian shape versus a uniform shape.  The error convolved posterior sampled SFHs of the two shapes look very similar, but the difference between their maximum likelihoods shows a preference for the Gaussian shape.  \citet{Bonaca_2020} fit a uniform SFH to a selection of GSE MSTOs from an earlier version of the H3 catalog, finding a truncation time $t_1=10.2~\pm~0.2$ Gyr while we instead find a time of $t_1=9.4~\pm~0.2$ Gyr.  This discrepancy might be explained by our different abundance selections;  \citet{Bonaca_2020} restricts \fehi$<-0.6$ while we instead find GSE stars up to \fehi$\approx-0.4$.  Given GSE's age---metallicity relation in Figure \ref{fig:n amrs} it is not surprising that our sample therefore also extends to slightly younger ages.  The sharp cut off in star counts younger than $9.4$ Gyr in Figure \ref{fig:n age hists} supports that these young stars as still legitimate members of GSE as opposed to contamination from another population.

\begin{figure*}
\epsscale{0.8}
\plotone{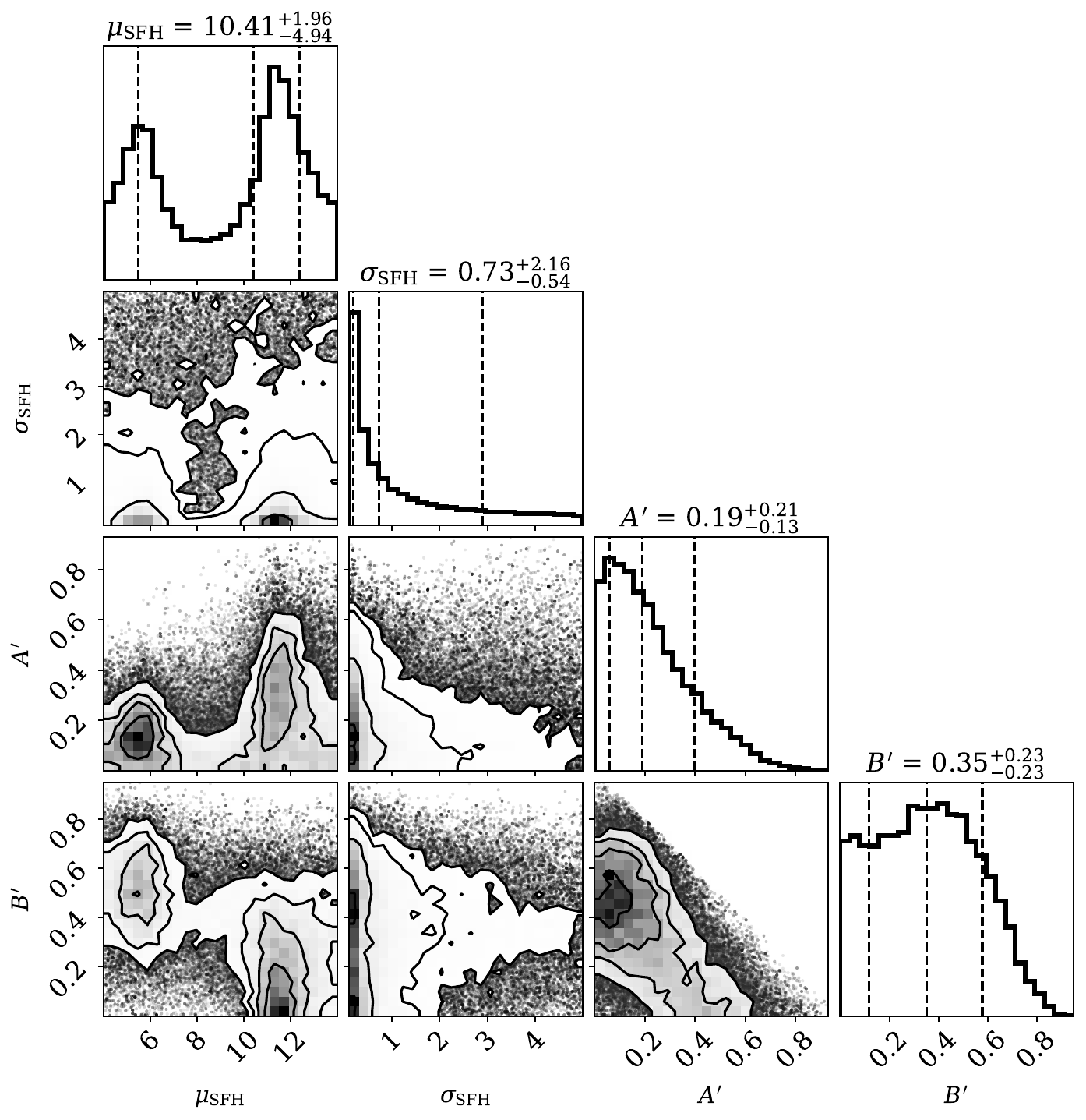}
\caption{Corner plot of Arjuna's SFH fit when including the SFH of GSE as an additional component, weighted by its own membership fraction $B^\prime$.  The bimodal results shows that Arjuna does not remain distinct when fitted with this additional GSE component.  One mode has a low value of $B^\prime$ and a high value of $A^\prime$ at approximately the same $\mu_\mathrm{SFH}$ as the fiducial fit.  But a significant second mode has a high value of $B^\prime$ with a low $A^\prime$ and a very young $\mu_\mathrm{SFH}$.  In this second mode the majority of Arjuna's age distribution is sufficiently explained by GSE's SFH.  In other words, Arjuna is not only indistinct from GSE in abundance space but is also indistinct in age space, further supporting a common origin.\label{fig:Arjuna_GSEcontamination}}
\end{figure*}

Figure \ref{fig:Arjuna_GSEcontamination} shows the corner plot of the SFH  posterior when we add an additional component for GSE in Arjuna's SFH fit (see Section \ref{section:results}).  The posterior is bimodal in $\mu_\mathrm{SFH}$ and most importantly, the membership fraction for a distinct Arjuna component $A'$ is actually less than the membership fraction for the GSE component $B'$.  In other words, GSE's star formation history is sufficient to explain the age distribution of Arjuna's MSTOs.  In fact when comparing to Figure \ref{fig:n age hists} we can see that in these cases where $B'>A'$, the primary SFH term $A'\psi_\mathrm{prim}$ is only capturing the three young stars with ages $\approx5-6$ Gyr.  The rest of the stars are consistent with the age distribution and SFH of GSE.  \citet{Naidu_2021} demonstrated that Arjuna can be plausibly connected to GSE in dynamical space, and the two are already indistinguishable in abundances \citep{Naidu_2020}.  Here we now see that the two are also indistinguishable in age space, further supporting that Arjuna does not have a distinct origin and is instead simply the high energy retrograde debris of GSE.

\begin{figure*}
\epsscale{1.0}
\plotone{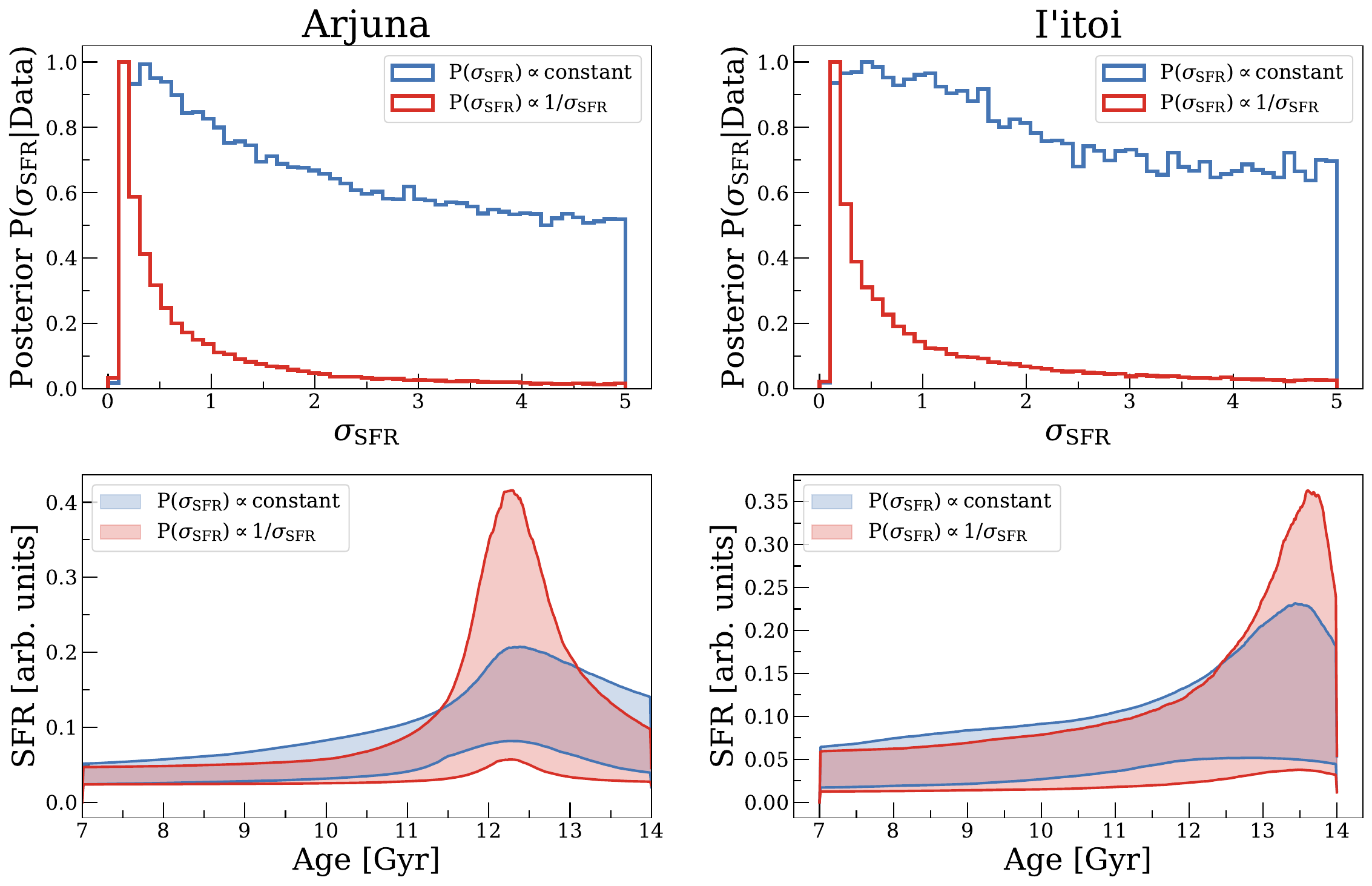}
\caption{Arjuna's results are plotted in the left column, and I'itoi's in the right column.  Top: Marginalized posteriors for the parameter $\sigma_\mathrm{SFH}$ when using either the uniform or scale invariant prior on $\sigma_\mathrm{SFH}$.    The modes of the marginalized posteriors using the uniform prior are at small values near $\sigma_\mathrm{SFH}\approx0.5-1.0$, but the fits were still largely uninformative so the tails of the marginal posterior are highly sensitive to the bounds of the uniform prior.  The scale invariant prior circumvents this issue and concentrates the posterior to smaller values, consistent with the mode when using the uniform prior. Bottom: The 16th to 84th percentile range of sampling from the full SFH posteriors under the uniform and scale invariant priors.  The scale invariant prior keeps the SFH width more compact, which is more in line with what we expect if Arjuna and I'itoi represent less massive progenitors.\label{fig:ArjunaIitoi_2priors}}
\end{figure*}

As mentioned in Section \ref{subsec:sfh fitting}, the SFH fits of Arjuna and I'itoi were uninformative on $\sigma_\mathrm{SFH}$, leaving the marginalized posterior of this parameter to be prior dominated.  This is an issue because the entire SFH posterior then depends on the arbitrary prior bounds.  For these two substructures we therefore resort to using a scale invariant prior of the form $P(\sigma_\mathrm{SFH})\propto1/\sigma_\mathrm{SFH}$.  Figure \ref{fig:ArjunaIitoi_2priors} shows the results when using this prior compared to using the default uniform prior on $\sigma_\mathrm{SFH}$.  The scale invariant prior keeps the width of the SFH more compact, which is more consistent with what we expect for less massive systems.  Of course the resulting SFH distributions in these cases is heavily conditioned on the choice of prior, so we are limited in the conclusions we can draw about the histories of Arjuna and I'itoi.

\bibliographystyle{aasjournal}

\end{CJK*}
\end{document}